\documentclass[12pt,tightenlines,
    onecolumn,
    preprintnumbers,
    amsmath,
    amssymb,
    preprint,
    nofootinbib,
    showpacs
]{revtex4}
\usepackage{color}
\usepackage{graphicx}
\usepackage{amsmath}
\usepackage{amssymb}
\usepackage{bm}
\usepackage{enumerate}

\newcommand{\nn}{\nonumber}

\newcommand{\E}{\mathcal{E}}
\begin{document}
\begin{titlepage}
\vspace{.5in}
\begin{flushright}

\end{flushright}
\vspace{0.5cm}

\title{Geodesic motions in extraordinary string geometry}

\author{Bogeun Gwak}
\email{rasenis@sogang.ac.kr}
\author{Bum-Hoon Lee}
\email{bhl@sogang.ac.kr}
\author{Wonwoo Lee}
\email{warrior@sogang.ac.kr} \affiliation{Department of Physics
and BK21 Division, and Center for Quantum Spacetime, Sogang
University, Seoul 121-742, Korea}
\author{Hyeong-Chan Kim}
\email{hckim@cjnu.ac.kr}
\affiliation{School of Liberal Arts and Sciences, Chungju National University, Chungju 380-702,
  Korea}

\begin{abstract}

{The geodesic properties of the extraordinary vacuum string
solution in (4+1) dimensions are analyzed by using Hamilton-Jacobi
method. The geodesic motions show distinct properties from those
of the static one.
Especially, any freely falling particle can not arrive at the horizon or singularity.
There exist stable null circular orbits and bouncing timelike and null geodesics.
To get into the horizon {or singularity}, a particle need to follow a non-geodesic trajectory.
We also analyze the orbit precession to show that the precession angle has distinct features for each geometry such as naked singularity, black string, and wormhole.
 }
\end{abstract}
\maketitle

\end{titlepage}

\section{Introduction}

A well known static vacuum hypercylindrical solution of the
Einstein equation in five dimensions is the Schwarzschild
blackstring solution, which is characterized by the ADM mass density
$M$ and has a horizon with topology $S^2\times R^1$. The
Schwarzschild black string background is known to be unstable
under small gravitational perturbations along the fifth direction
if the corresponding length is greater than some threshold - the
so-called Gregory-Laflamme (GL) instability~\cite{GL,after}. The
full non-linear evolution of a black string beyond this threshold
might result in a black string breaking up into separate black
holes or would settle into a stable, static non-uniform black
string state~\cite{Horowitz}.

Because of the instability, the hypercylindrical vacuum
solutions~\cite{Kramer,chodos} of the Einstein equation in five
dimensions, which has two independent parameters, was
reinvestigated by Lee~\cite{lee} in which the physical implication
of the two parameters were correctly interpreted as ADM mass and
tension densities~\cite{ADM}. It was also pointed out that the well-known
Schwarzschild black string solution corresponds to the case that
the tension-to-mass ratio is exactly one half. This {implies} that the
Schwarzschild black string, which was believed to be characterized by the mass density only, is indeed a special case of a wider class of solutions characterized by
tension density as well. In Ref.~\cite{kang} the geometric
properties of this class of spacetime with arbitrary tensions
were investigated in detail.
In Ref.~\cite{yun} the author studied dyonic brane solutions with tension.

Note also that in $5$-dimensional spacetime, there is another
class of stationary solutions~\cite{kimjl} characterized by the ADM mass and tension densities and momentum along the fifth coordinate.
It was shown that the tension density can be made to be the same as the mass density by using the boost symmetry of an asymptotic observer.
In that case, the ADM momentum takes its minimum (non-zero) size which we call ``momentum charge" in this paper.
Therefore, it is impossible to change the ``momentum charge" by
using the boost transformation of asymptotic observer.
Let $\mathfrak{p}$ denotes the ``momentum
charge"-to-mass ratio. As $\mathfrak{p}$ increases the metric
behaves as a spacetime with a naked singularity, a black
string surrounded by an event horizon (or null singularity), or a
wormhole. We summarize some of the main properties of the solution
in the subsection~\ref{Sec:1A}.

Geodesics describing the motion of massive particles or light rays
in a given spacetime is one of
the best methods to illustrate the geometry of spacetime when the
effect of backreaction is negligibly small.
In the Schwarzschild
spacetime, stable circular orbits for massive
particles exist at $r \geq 6G_4M$ only. The orbit at $r=6G_4M$ is
called the marginally stable circular orbit. The circular orbits for $r < 6G_4M$
are unstable. On the other hand, there are no
stable circular orbits for photon in the spacetime. The light ray
with impact parameter smaller than specific value is captured by
the strong gravitational field. The value of critical impact
parameter $b$ is $3\sqrt{3} G_4M$ in the Schwarzschild case. The
capture cross-section for light rays is $\pi b^2=27\pi G_4^2M^2$,
whereas the geometric cross-section is $\pi r_s^2 = 4\pi
G_4^2M^2$~\cite{Misner}. These are distinguished from Newtonian
features of the strong gravitational field. Geodesic motions for a
5-dimensional magnetized Schwarzschild-like solution were studied
in Ref.~\cite{matos}, the behavior of a test particle moving in
the spacetime is interpreted as a massive magnetic dipole coupled
with a massless scalar field. In Ref.~\cite{gll}, the authors
studied the geodesic motions of a massive particle and a light ray
in the hyperplane orthogonal to the symmetry axis in the
5-dimensional hypercylindrical spacetime~\cite{lee}. The
diffraction angle for bending light and some properties of orbits
are analyzed by using analytic and numerical methods. In this
paper we investigate how much these features of the geodesics
change in the presence of a ``momentum charge".

In Sec. II, we briefly summarize the stationary metric of the
extra-ordinary blackhole and find the geodesic motions by
using the Hamilton-Jacobi method. In Sec. III, we study the radial and angular
motions of the geodesics by using the effective potential method.
Sec. IV is devoted for summary and discussions.

\section{Geodesic motions in extra-ordinary solutions}
The most general form of the metric for the spherically
symmetric stationary spacetime with a translational symmetry along the
fifth spatial coordinate in five dimensions is
$$
ds^2 = -F (dt- X dz)^2 +G[d\rho^2+\rho^2(d\theta^2+\sin^2\theta d\phi^2)]+ H dz^2,
$$
where $F$, $G$, $H$, and $X$ are functions of the ``isotropic"
radial coordinate $\rho$ only. Note that the fifth direction is
not assumed to be flat in general, {\it i.e.}, $H\neq 1$. Such a
stationary spacetime was considered in
Refs.~\cite{jlkim,velocityFD}. Time-dependent solution in a
separable form were found in Ref.~\cite{Liu}. A class of solutions
allowing the $z$-dependence was also considered in
Ref.~\cite{Billyard}.

The stationary vacuum hypercylindrical solutions of Einstein
equation in 5-dimensions are classified by three classes. First is
static and is characterized by ADM ``mass'' and ``tension" densities~\cite{lee}.
Second is stationary and is characterized by ADM ``mass" and
``momentum charge" densities. Third consists of wormhole like
solutions. The static solution is well analyzed in
Ref.~\cite{kang} and its geodesic motion was given in
Ref.~\cite{gll}. The third solution was known in
Ref.~\cite{chodos}. The authors in Ref.~\cite{gll} studied the
geodesic motions of a massive particle and a light ray in the
hyperplane orthogonal to the symmetry axis.
The behaviors of the null geodesics were classified into two categories by the values of the tension-to-mass ratio $a$.
The deflection angles are qualitatively similar to those of the black
hole for $-1 \leq a \leq 2$ and different for $a < -1$ or $a > 2$.
There exists a parameter range in which the
singularity is a weakly naked one, $-1 < a < 2$ and a strongly
naked one, $a \leq -1$ or $a \geq 2$, in that analysis. For the
timelike geodesics, there is a stable circular orbits for $a <
-1$. One of the characteristics of the timelike case is that there
exists a marginally stable circular orbit for $-1 < a < 2$. The
angular momentum and the radius of this marginal orbit were
numerically obtained. As a result, the behavior of trajectories
has similar properties to that of Schwarzschild black hole if $-1
\leq a \leq 2$. Thus the solution with $-1 \leq a \leq 2$ corresponds
to the weakly naked one following the definition of Virbhadra and
Ellis~\cite{ve}.

A simplest nontrivial stationary hypercylindrical solution with momentum
along the string direction is the boosted
Schwarzschild and Kaluza-Klein bubble solutions with compact
extra-dimension~\cite{velocityFD}. It was shown that the momentum
parameter governs a global property of the spacetime (so the
spacetime is locally indistinguishable to that of the static one)
and there remains nontrivial frame dragging effect. The general
stationary solution was first given in Ref.~\cite{chodos} and was
known  in Ref.~\cite{kimjl} to be characterized by two parameters,
the ``mass" and ``momentum charge" densities.

\subsection{Summary of the extra-ordinary solution}\label{Sec:1A}
The ``momentum charge" density $\cal P$ is a new characteristic
quantity defined by the minimum value of ADM momenta seen by
asymptotic observers. The extra-ordinary string solution is a
class of solutions of Einstein equation in $5$-dimensions which
has a non-vanishing ``momentum charge" density in addition to the
``mass" density~\cite{kimjl}. We define the ratio of the
``momentum charge"-to-``mass",
\begin{equation}\label{p}
\mathfrak{p} \equiv \frac {\cal P}{M} ,
\end{equation}
where $M$ is the ADM mass density. The metric of the
vacuum stationary hypercylindrical solution of Einstein equation
in 4+1 dimensions is
\begin{eqnarray}
ds^2&=& \left(\frac{1+K/\rho}{1-K/\rho}\right)^{-\frac{2}{\sqrt{3}\sqrt{1-3\mathfrak{p}^2}}} \left[-(\omega^t)^2+(\omega^z)^2\right] \\
 &+&
\left(\frac{1+K/\rho}{1-K/\rho}\right)^\frac{4}{\sqrt{3(1-3\mathfrak{p}^2)}}
\left(1-\frac{K^2}{\rho^2}\right)^2 (d\rho^2+\rho^2
d\Omega_{(2)}).
 \nn
\end{eqnarray}
The time-like 1-form $\omega^t$ and the space-like 1-form
$\omega^z$ are
\begin{eqnarray} \label{eq:D}
\omega^t = \cos\Upsilon \, dt - \sin\Upsilon \,dz, \quad \omega^z
= \sin\Upsilon \,dt +\cos\Upsilon \, dz,
\end{eqnarray}
where $t$ and $z$ are Killing coordinates. The argument $\Upsilon$
is a function of radial coordinate $\rho$ given by
\begin{eqnarray} \label{upsilon}
\Upsilon(\rho)=\frac{\sqrt{3}\mathfrak{p}}{\sqrt{1-3\mathfrak{p}^2}}
\,\log\frac{1+K/\rho}{1-K/\rho}, \quad
\end{eqnarray}
which monotonically increases from zero to infinity as $\rho$
decreases from infinity to $K$ for positive $\mathfrak{p}$.

The ADM tension along the fifth direction is the same as the mass
density and the parameter $K$ is given by
\begin{eqnarray}
K= \frac{G_5 M\sqrt{1-3\mathfrak{p}^2}}{\sqrt{3}}.
\end{eqnarray}
The ADM momentum of this metric is the same as the ``momentum charge". The ADM momentum, $P$, with respect to a moving observer with
velocity $v$ along the $z$ direction satisfies $P/{\cal P}=
\sqrt{1+4q^2}\geq 1$ where the boost parameter satisfies
$\frac{q}{\sqrt{q^2+1/4}}=\frac{2v}{1-v^2}$.
The reality of the parameter $K$ restricts the value of
$\mathfrak{p}$ to $-\frac1{\sqrt{3}}\leq \mathfrak{p}\leq
\frac1{\sqrt{3}}$. Note that the metric is invariant under
the change $\mathfrak{p}\to -\mathfrak{p}$ and $z\to -z$.
Therefore, we restrict ourselves to the case with non-negative
$\mathfrak{p}$ without loss of generality. Now we summarize some
of the main properties of the metric.

\begin{itemize}
\item The ``momentum charge" density is restricted to the range $|
\mathfrak{p}|\leq \frac{1}{\sqrt{3}}$.

\item There is a curvature singularity at $\rho=K$ unless
$|\mathfrak{p}|=\sqrt{\frac{5}{27}}$ or $ \frac{\sqrt{2}}{3}\leq
|\mathfrak{p}| \leq \frac{1}{\sqrt{3}}$.
    The curvature singularity is shown to be naked if $|\mathfrak{p}|< \frac1{2\sqrt{3}}$ or null if $\frac1{2\sqrt{3}}\leq |\mathfrak{p}| < \frac{\sqrt{2}}{3}$.

\item  The $\rho=K$ surface lies at spatial infinity for
$\frac{\sqrt{2}}{3}\leq |\mathfrak{p}| \leq \frac{1}{\sqrt{3}}$
and at a finite distance else.
    Especially, the spacetime becomes asymptotically flat as $\rho \to K$ for $\frac{\sqrt{2}}{3}<|\mathfrak{p}| \leq \frac{1}{\sqrt{3}} $.

\item The zero ``momentum charge" solution is the static solution with $M=\tau$
in Ref.~\cite{lee}.

\item The metric for $\frac{\sqrt{2}}{3}\leq |\mathfrak{p}| \leq
\frac{1}{\sqrt{3}}$ describes a spacetime with  a wormhole-like
geometry.

\item The area of the $S^2$ sphere decreases to some finite value
and then bounces back to infinity as one approaches to the
$\rho=K$ surface from infinity.

\item On the other hand, the proper length $L$ of a spacelike
segment in $t$-$z$ plane shrinks down to zero as one approaches to
the $\rho=K$ surface.

\item As a result of the competition of the $S^2$ area and the
segment length, the total area of the segment $S^2\times L$ of the
$\rho=K$ surface turns out to vanishes for $|\mathfrak{p}|<
\frac1{2\sqrt{3}}$, becomes a finite number for $|\mathfrak{p}|=
\frac1{2\sqrt{3}}$, and diverges for  $|\mathfrak{p}|>
\frac1{2\sqrt{3}}$.

\item There appears an extremely strong velocity frame dragging.
As a result, the coordinates $t$ and $z$ exchange their roles as a
time and as a space successively whenever the radial coordinate
$\Upsilon$ is increased by $\frac{\pi}2$.
The Killing vectors $(\partial/\partial t)_a$
[$(\partial/\partial z)_a$] is timelike [spacelike]
asymptotically. However, as $\rho$ decreases, it successively
becomes spacelike [timelike] then timelike [spacelike] and so on.
This account for why the peculiar conserved ``momentum charge"
cannot be gauged away by the motion of an asymptotic observer.
\end{itemize}

\subsection{Geodesic motion by using Hamilton-Jacobi method}

Since we are considering the geodesic motions, we set the
zenith angle $\theta=\pi/2$ without loss of generality. The metric
has a spherical symmetry and the translational symmetries along
$z$ and $t$. Therefore, we have three Killing vectors
$(\partial_t)^a$, $(\partial_z)^a$, and $(\partial_\phi)^a$ and
the corresponding conserved quantities the energy $E$, the
momentum along the string direction $p_z$, and the angular
momentum $L$.

To have the action of a geodesic for a particle, we turn our
attention to the Hamilton-Jacobi method, which is useful to obtain
the action itself~\cite{Misner}. The classical action, $S$, is
taken to satisfy the relativistic Hamilton-Jacobi equation
$$
g^{ab} \partial_a S \partial_b S +m^2=0.
$$
We renormalize $m=1$ for timelike geodesics and $m=0$ for
lightlike geodesics, then we have
\begin{eqnarray}\label{HJ}
0&=& m^2+\frac{\cos2\Upsilon\left[-(\frac{\partial S}{\partial t})^2+(\frac{\partial S}{\partial z})^2\right]
    +2\sin2\Upsilon \frac{\partial S}{\partial t}\frac{\partial S}{\partial z}}{D^{-\frac{2}{\sqrt{3(1-3\mathfrak{p}^2)}}} }
    +\frac{(\frac{\partial S}{\partial \rho})^2}{G}+\frac{(\frac{\partial S}{\partial \theta})^2}{G\rho^2}+\frac{(\frac{\partial S}{\partial \phi})^2}{G\rho^2\sin^2\theta},
\end{eqnarray}
where
\begin{eqnarray}
G = g_{\rho\rho}=
\left(1-\frac{K^2}{\rho^2}\right)^2\left(\frac{1+K/\rho}{1-K/\rho}\right)^{\frac{4}{\sqrt{3(1-3\mathfrak{p}^2)}}},
\quad D = \frac{1+K/\rho}{1-K/\rho} .
\end{eqnarray}

Since we have three Killing vectors, those symmetries generate
three conserved quantities in the sense that $S$ must be a
covariant constant under its action: $\mathfrak{L}_{k} = k^a
\partial_a S = \varepsilon$ where $k^a$ is a Killing vector and
$\varepsilon$ is the corresponding conserved constant. The
physical interpretation of this is that $\partial_a S$ is
cotangent to a geodesic and $\varepsilon$ is a conserved quantity
along that geodesic. Now, the solution can be obtained to be
\begin{eqnarray}
S= -E t+p_z z+L \phi+ S_\frac{\pi}2(\rho),
\end{eqnarray}
where $E$, $p_z$, and $L$ represent the energy, momentum along $z$
direction, and the angular momentum, the subscript $\frac{\pi}2$
implies that the zenith angle $\theta=\pi/2$ and the $\rho$ dependent
part is
\begin{eqnarray}
S_\frac{\pi}2(\rho)&=& \E \int^\rho d\rho \sqrt{ G
D^{\frac{2}{\sqrt{3(1-3\mathfrak{p}^2)}}}\cos 2(\alpha+\Upsilon)
    - \frac{m^2 G}{\E^2}-\frac{L^2}{\E^2 \rho^2}}.
\end{eqnarray}
In this equation we introduce new parameters,
\begin{equation}
\tan \alpha =\frac{ p_z}{E}, \quad \E=\sqrt{p_z^2+E^2}.
\label{eqphi}
\end{equation}

The motion of the coordinates $t$, $z$, and $\phi$ can be obtained
by using variation of $S$ with respect to $E$, $p_z$, and $L$:
\begin{eqnarray} \label{t}
t &=& \int^\rho d\rho \frac{GD^{\frac{2}{\sqrt{3}
\sqrt{1-3\mathfrak{p}^2}}} \cos (2\Upsilon+\alpha)}{
    \sqrt{G D^{\frac{2}{\sqrt{3} \sqrt{1-3\mathfrak{p}^2}}}\cos 2(\Upsilon+\alpha)- \frac{m^2 G}{\E^2}-\frac{L^2}{\E^2\rho^2} } }
    , \\  \label{z}
z&=&  \int^\rho d\rho \frac{GD^{\frac{2}{\sqrt{3}
\sqrt{1-3\mathfrak{p}^2}}} \sin (2\Upsilon+\alpha)}{
    \sqrt{G D^{\frac{2}{\sqrt{3} \sqrt{1-3\mathfrak{p}^2}}}\cos 2(\Upsilon+\alpha)- \frac{m^2 G}{\E^2}-\frac{L^2}{\E^2\rho^2} } }
   , \\    \label{theta}
\phi&=&\frac{L}{\E} \int^\rho \frac{d\rho}{\rho^2} \frac{1}{
    \sqrt{ G D^{\frac{2}{\sqrt{3} \sqrt{1-3\mathfrak{p}^2}}}\cos 2(\Upsilon+\alpha)- \frac{m^2 G}{\E^2}-\frac{L^2}{\E^2\rho^2} } }.
\end{eqnarray}
The geodesics are well defined only in the region where the
argument of the square root is positive definite.

We introduce a new time coordinate $t'$ and a new space
coordinate $z'$ by
\begin{eqnarray} \label{t'}
t'\equiv t\cos\Upsilon+z\sin\Upsilon
 &=&  \int^\rho d\rho \frac{GD^{\frac{2}{\sqrt{3} \sqrt{1-3\mathfrak{p}^2}}} \cos(\alpha+ \Upsilon)}{
    \sqrt{G D^{\frac{2}{\sqrt{3} \sqrt{1-3\mathfrak{p}^2}}}\cos 2(\Upsilon+\alpha)- \frac{m^2 G}{\E^2}-\frac{L^2}{\E^2\rho^2} } }, \\
z'\equiv -t\sin\Upsilon+z\cos\Upsilon &=&\int^\rho d\rho
\frac{GD^{\frac{2}{\sqrt{3} \sqrt{1-3\mathfrak{p}^2}}} \sin
(\Upsilon+\alpha)}{
    \sqrt{G D^{\frac{2}{\sqrt{3} \sqrt{1-3\mathfrak{p}^2}}}\cos 2(\Upsilon+\alpha)- \frac{m^2 G}{\E^2}-\frac{L^2}{\E^2\rho^2} }  }.  \nn
\end{eqnarray}
Note that the vector $\left(\frac{\partial}{\partial t'}\right)^\mu$ is everywhere timelike contrary to $\left(\frac{\partial}{\partial t}\right)^\mu$
and $t'=t$ asymptotically. By using the time $t'$, the action can
be rewritten as
\begin{eqnarray}
S &=& \E\left[\cos(\Upsilon+\alpha)\, t'+\sin(\Upsilon+\alpha) \,z'\right] +L\theta \\
&&+ \E \int^{\rho(t')} \frac{dt'}{\cos(\alpha+\Upsilon)}
    \left[\cos2(\alpha+\Upsilon) -\frac{m^2}{\E^2D^{\frac{2}{\sqrt{3} \sqrt{1-3\mathfrak{p}^2}}}}-\frac{L^2}{\E^2 G D^{\frac{2}{\sqrt{3} \sqrt{1-3\mathfrak{p}^2}}}\rho^2}  \right]. \nn
\end{eqnarray}

How much time will it take if a rocket travels from $K+\epsilon$
to $\rho$ through radial null (non)-geodesic motion? The outgoing
velocity of geodesics with respect to the time coordinate $t'$ is
evaluated by differentiating Eq.~(\ref{t'}). Of all outgoing
radial null geodesics at $\rho$, the maximal velocity is achieved
by taking $\alpha=-\Upsilon(\rho) $:
\begin{eqnarray}
v_{\rm max} &=& \left(\frac{d\rho}{dt'}\right)_{\rm
max}=\frac1{G^{1/2}D^{\frac{1}{\sqrt{3}
\sqrt{1-3\mathfrak{p}^2}}}}=
\left(1-\frac{K}{\rho}\right)^{\frac{\sqrt{3}}{\sqrt{1-3\mathfrak{p}^2}}-1}
\left(1+\frac
K{\rho}\right)^{-1-\frac{\sqrt{3}}{\sqrt{1-3\mathfrak{p}^2}}} \nn
.
\end{eqnarray}
Therefore, we adjust the energy and momentum of the rocket to
satisfy $\alpha= -\Upsilon(\rho)$ at a given time. The shortest
time for one information traveling from $K+\epsilon$ to $\rho$
becomes
\begin{eqnarray}
\Delta t'= \int_{K+\epsilon}^\rho
\left(1-\frac{K}{\rho'}\right)^{1-\frac{\sqrt{3}}{\sqrt{1-3\mathfrak{p}^2}}}
\left(1+\frac
K{\rho'}\right)^{1+\frac{\sqrt{3}}{\sqrt{1-3\mathfrak{p}^2}}}d
\rho'.
\end{eqnarray}
For small $\epsilon$, we get the explicit values:
\begin{eqnarray}
\Delta t'=\left\{ \begin{tabular}{ll}
                $ \frac{2^{1+\frac{\sqrt{3}}{\sqrt{1-3\mathfrak{p}^2}}} K}{2-\frac{\sqrt{3}}{\sqrt{1-3\mathfrak{p}^2}}}\left(\frac \epsilon K \right)^{
                2-\frac{\sqrt{3}}{\sqrt{1-3\mathfrak{p}^2}}}+$finite terms,~~& $\sqrt{1-3\mathfrak{p}^2} \neq \frac{\sqrt{3}}{2}$, \\
                $ -8 K\log\left(\frac \epsilon K \right)+$finite
                terms,
                & $\sqrt{1-3\mathfrak{p}^2} = \frac{\sqrt{3}}{2}$. \\
                \end{tabular} \right.
\end{eqnarray}
In the $\epsilon \to 0$ limit, this value diverges for
$|\mathfrak{p}|\geq \frac{1}{2\sqrt{3}}$, which signals the
presence of an event horizon or a null singularity at $\rho=K$.

\section{Effective potential for the radial motions}
The momentum is given by $p^\rho =
\frac{dx^\rho}{d\lambda}=\frac{d \rho}{d\lambda}$ where $\lambda$
is the affine parameter. The effective potential for the radial
motion of a geodesic can be obtained by inserting the conserved
quantities $E$, $p_z$, and $L$ into the Hamilton-Jacobi
equation~(\ref{HJ}):
\begin{eqnarray}
\left(\frac{d\rho}{d\lambda}\right)^2&=& (g^{\rho\rho} p_\rho)^2
=\frac{1}{G^2}\left(\frac{\partial S}{\partial \rho}\right)^2=-
V_{eff}(\rho),
\end{eqnarray}
where the effective potential for a particle with mass $m$ and
three conserved quantities $(E,p_z,L )$ becomes
\begin{eqnarray} \label{Veff}
V_{eff}(\rho)
&=&\frac{(1-K/\rho)^{\frac{2}{\sqrt{3(1-3\mathfrak{p}^2)}}-2} }{
    (1+K/\rho)^{\frac{2}{\sqrt{3(1-3\mathfrak{p}^2)}}+2} }\Big[- \E^2\cos 2(\alpha +\Upsilon)  \nn  \\
& &+ m^2
\left(\frac{1-K/\rho}{1+K/\rho}\right)^{\frac{2}{\sqrt{3(1-3\mathfrak{p}^2)}}}
    +\frac{L^2}{\rho^2}\frac{(1-K/\rho)^{\frac{6}{\sqrt{3(1-3\mathfrak{p}^2)}}-2} }{
    (1+K/\rho)^{\frac{6}{\sqrt{3(1-3\mathfrak{p}^2)}}+2} }\Big] .
\end{eqnarray}
\begin{figure}[htbp]
\begin{center}
\includegraphics[width=.6\linewidth,origin=tl]{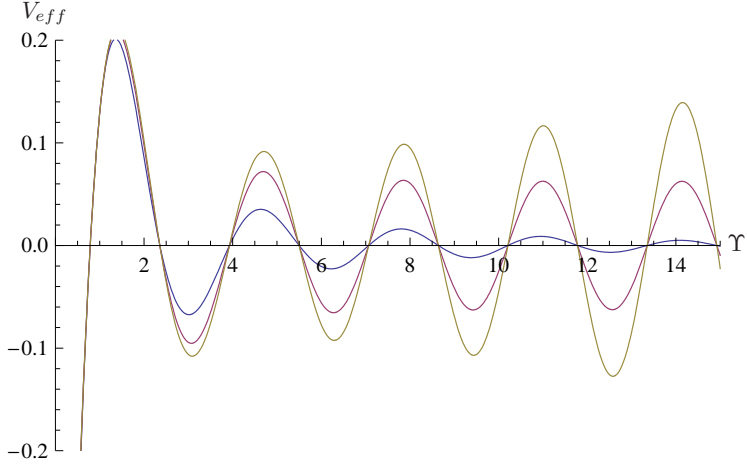}
\end{center}
\caption{(color online) Effective potential $V_{eff}(\Upsilon)$
for null radial geodesics. Here, we choose $\E=1$ and $\alpha=0$.
The blue, purple, and yellow curves correspond to the cases
$\mathfrak{p}\simeq 0.483$, $\mathfrak{p}=\sqrt2/3$, and
$\mathfrak{p} \simeq 0.365$ from the horizontal axis,
respectively. For values $\alpha\neq 0$, the roots of the curves
move horizontally. The origin corresponds to $\rho\to \infty$ and
the $\Upsilon\to \infty$ corresponds to $\rho=K$.
 } \label{fig:fig1}
\end{figure}

Now let us analyze the effective potential in detail. For
comparison, we put the effective potential of a moving particle
having two conserved quantities $(E,L)$ in a Schwarzschild black
hole spacetime:
$$
V_{\rm Schwarzschild}(r)= m^2-E^2-\frac{2m^2 M}{r}+\frac{L^2}{r^2}-\frac{2M L^2}{r^3},
$$
where $M$ is the mass of the Schwarzschild black hole and $r$ is
the arial radial coordinate. This potential will be negative at
$r\to \infty$ and increases with $r$ because of the mass term
[$(1/r)$ potential]. It has a local minimum since the angular
momentum gives rise to repulsive force effectively. For smaller
angular momentum than $2\sqrt{3}M m^2$, the local minimum
disappears and the potential does not allow any stable (and even
unstable) circular geodesic.

Now let us return to the present case. Note that for $\rho \sim
K$, the effective potential is dominated by the first term in
Eq.~(\ref{Veff}), which is independent of the mass or angular
momentum and oscillates indefinitely around zero. The amplitude of
the oscillation is governed by the exponent of $(1-K/\rho)$,
$$
\frac{2}{\sqrt{3(1-3\mathfrak{p}^2)}}-2.
$$
Therefore, as $\rho \to K$ the size of $V_{\rm eff}$ decreases to zero if $|\mathfrak{p}|
> \sqrt{2}/3$, increase to infinity if $|\mathfrak{p}|<\sqrt{2}/3$, and
approach to a constant value if $|\mathfrak{p}|=\sqrt{2}/3$. The
oscillatory nature of the effective potential, which comes from
the cosine function, is one of the main differences from that of
the ordinary static black hole or string solutions with vanishing ``momentum charge". In
the case of the Schwarzschild black hole, the energy squared
simply determines the height of the effective potential at $\rho\to \infty$.
However,
at the present case $E^2 + \mathfrak{p}^2_z$ determines the
amplitude of oscillation. Because of the oscillation, there
appear both the forbidden regions and allowed regions for a
particles with given conserved quantities $(E,p_z,L)$. In the case
of a null geodesics with zero angular momentum, the allowed regions
are specified to be
$$
 n\pi -\alpha  -\frac{\pi}4\leq \Upsilon \leq n\pi-\alpha  +\frac{\pi}4 \,.
$$

The asymptotic region will be allowed here if $-\frac{\pi}4 \leq
\alpha \leq \frac{\pi}4$ or $\frac{3\pi}4\leq \alpha \leq
\frac{5\pi}4$ and only the first region is relevant when we
restrict ourselves to the particles with positive energy,
satisfying $E\geq |p_z|$. Noting this, we may write the allowed
region including the asymptotic in terms of the radial coordinate
$\rho$,
\begin{eqnarray} \label{AllowedRegion}
&&\rho\geq \rho_{\rm boundary}, \\
&&
\coth\left[\frac{\sqrt{1-3\mathfrak{p}^2}}{2\sqrt{3}\mathfrak{p}}\left(n\pi-\alpha
+\frac{\pi}4\right)\right] \leq \frac{\rho}{K} \leq
\coth\left[\frac{\sqrt{1-3\mathfrak{p}^2}}{2\sqrt{3}\mathfrak{p}}\left(n\pi
-\alpha -\frac{\pi}4\right)\right]
   , \nn
\end{eqnarray}
where $n$ is any positive integer and the outermost boundary of
the allowed region $\rho_{\rm boundary}$ is
\begin{eqnarray}
\rho_{\rm boundary}(\alpha,L=0,m=0) =
\coth\left[\frac{\sqrt{1-3\mathfrak{p}^2}}{2\sqrt{3}\mathfrak{p}}\left(\frac{\pi}4-\alpha
\right)\right] .
\end{eqnarray}
For small value of momentum charge-to-mass ratio, $\mathfrak{p}\ll
1$, the boundary approaches to $K$ and essentially the boundary
becomes the $\rho=K$ surface at $\mathfrak{p}=0$. For
$\mathfrak{p}\sim 1/\sqrt{3}$ the position of the boundary becomes
large $\rho_{\rm boundary} \to \infty$. In general, the explicit
value of $\rho_{\rm boundary}(\alpha,L,m)$ will be dependent on
$L$ and $m$ explicitly. We compare the radial geodesic motion for
$\rho>\rho_{\rm boundary}$ with that of the Schwarzschild black
hole outside of the angular momentum barrier. We show that the
properties of the barrier are quite different from that of the
Schwarzschild in the following senses,
\begin{itemize}
\item Any geodesics cannot go over the potential barrier however
high energy the particle has.
    This fact is denoted by the presence of $\rho_{\rm boundary}$ for any values of $E$, $L$, and $p_z$.
\item There exists the boundary $\rho_{\rm boundary}$ for null
geodesics even in the absence of an angular momentum barrier.
Therefore, the origin of this barrier is not due to the angular
momentum but due to the ``momentum charge". \item The position of
the boundary is crucially dependent on the momentum-to-energy
ratio $p_z/E=\tan\alpha$ of the particle rather than the energy
itself.
\end{itemize}

The asymptotic region is not included in the
allowed region if $\frac{\pi}4 < \alpha < \frac{3\pi}4$ or
$\frac{5\pi}4< \alpha < \frac{7\pi}4$, where $|p_z| > E$. The
allowed region in terms of the radial coordinate $\rho$ in this
case is given by the second line of Eq.~(\ref{AllowedRegion}).
Note that all allowed regions are compact in $\rho$ and there is
no asymptotic region. Note also that the coordinate $t$ is not a
time in these regions. Rather, the coordinate $z$ does the role of
time, therefore we may understand the condition $|p_z|
> E$.

For large $\rho$, the effective potential takes the form:
\begin{eqnarray} \label{V:series}
&&V_{eff}(\rho)\sim  -E^2+p_z^2 +m^2-\frac{2m^2+p_z^2-E^2 - 6\mathfrak{p}p_z E}{3 }\,\frac{4G_5 M}{\rho} \\
&&
+\left\{\frac{L^2}{G_5^2M^2(\frac13-\mathfrak{p}^2)}-6\left(\mathfrak{p}^2-\frac{19}{9}\right)m^2
   -2\E^2\left[\left(\frac{7}{3}-15\mathfrak{p}^2\right)\cos2\alpha+8\mathfrak{p} \sin 2\alpha\right]\right\}\frac{G_5^2 M^2}{3\rho^2}
     .\nn
\end{eqnarray}
Observing the zeroth order term, the region $\rho\to \infty$ is
allowed to particles with $E^2\geq p_z^2+m^2$. The condition is related to the value of momentum along the extra direction, contrary to the case of the Schwarzschild.

Seeing the  $O(1/\rho)$ term, the potential increases with $\rho$ if
$$
0\leq E<-3\mathfrak{p} p_z+\sqrt{(9\mathfrak{p}^2+1)p_z^2+2m^2} ,
$$
where we set the lower bound of $E$ to zero since we deal with the situation around $\rho\to \infty$.
We summarize various properties of the effective potential:
\begin{itemize}
\item In the absence of ``momentum charge" ($\mathfrak{p}=0$), the
asymptotic form of the effective potential increases with $\rho$
if $p_\perp < m$, where $p_\perp= \sqrt{E^2-p_z^2-m^2}$ is the
momentum along the $3$-dimensional space other than the
extra-dimension. For particles with high momentum, $p_\perp >  m$,
the effective potential may decreases as $\rho$ increases, which
behavior does not happen for the Schwarzschild geometry. In the
case of a null geodesics ($m=0$), the effective potential decreases with $\rho$.

\item For null geodesics moving along the opposite direction to
the dragging ($\mathfrak{p}p_z<0$), the asymptotic form of the
effective potential increases with $\rho$ if $p_\perp <
-3\sqrt{2}\mathfrak{p}p_z (1+\sqrt{1+1/(9\mathfrak{p}^2)})^{1/2}$.
This implies that there exist a stable circular orbit for a light
at some $\rho> \rho_{\rm boundary}$, which is completely different
from the cases of  $\mathfrak{p}=0$ and the Schwarzschild black
hole. Some example of this behavior are shown in the next section.
For other cases except for this, the effective potential
decreases asymptotically for the null geodesics.

\item  For timelike geodesics, the asymptotic form of the
effective  potential can decrease or increase according to the
parameter.
    When it decreases asymptotically, the effective potential does not have any local minimum in the outside  $(\rho>\rho_{\rm boundary})$ of the classically forbidden region.
\end{itemize}

\section{The geodesic motions }

\begin{figure}[t]
\begin{center}
\includegraphics[width=1.3in]{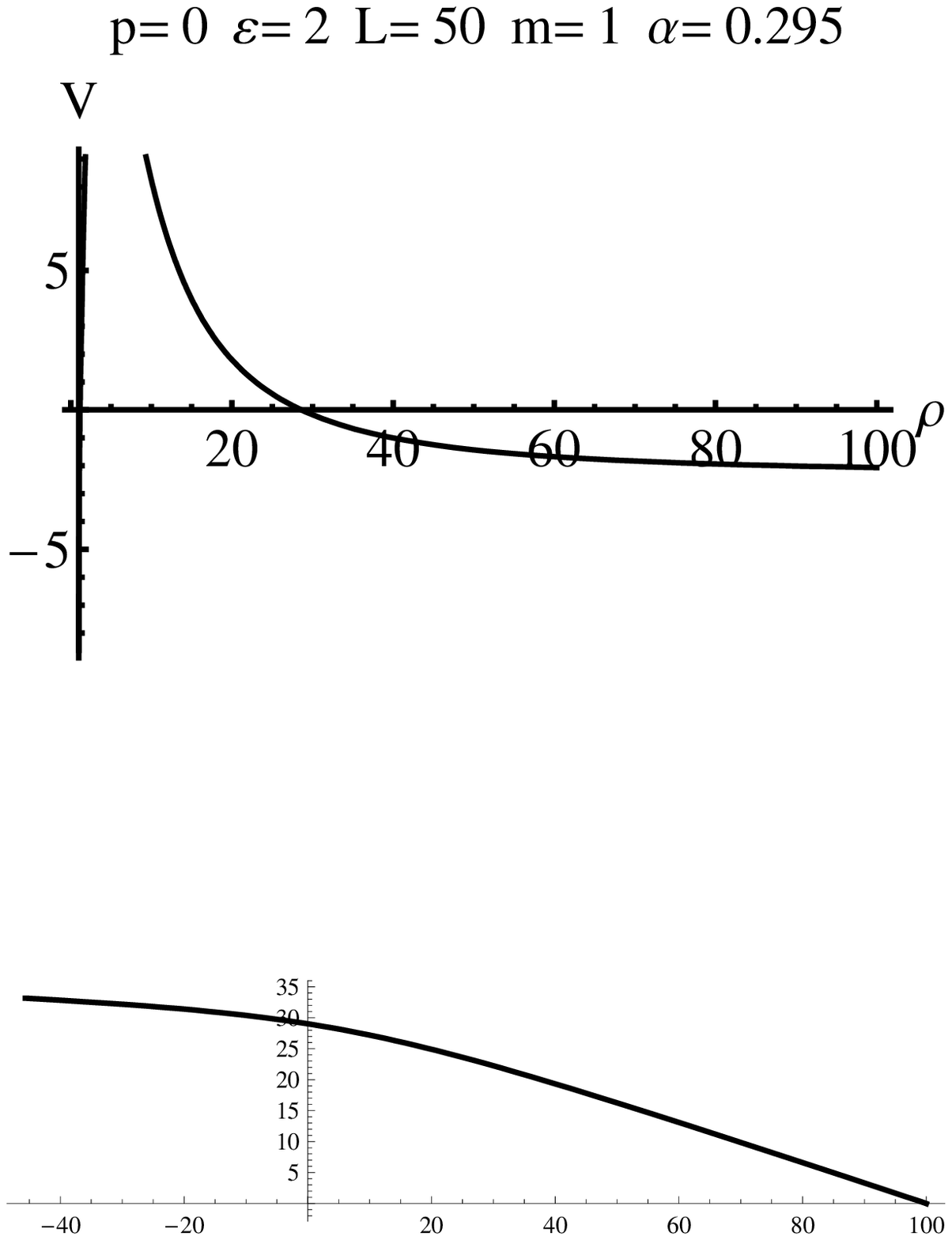}
\includegraphics[width=1.3in]{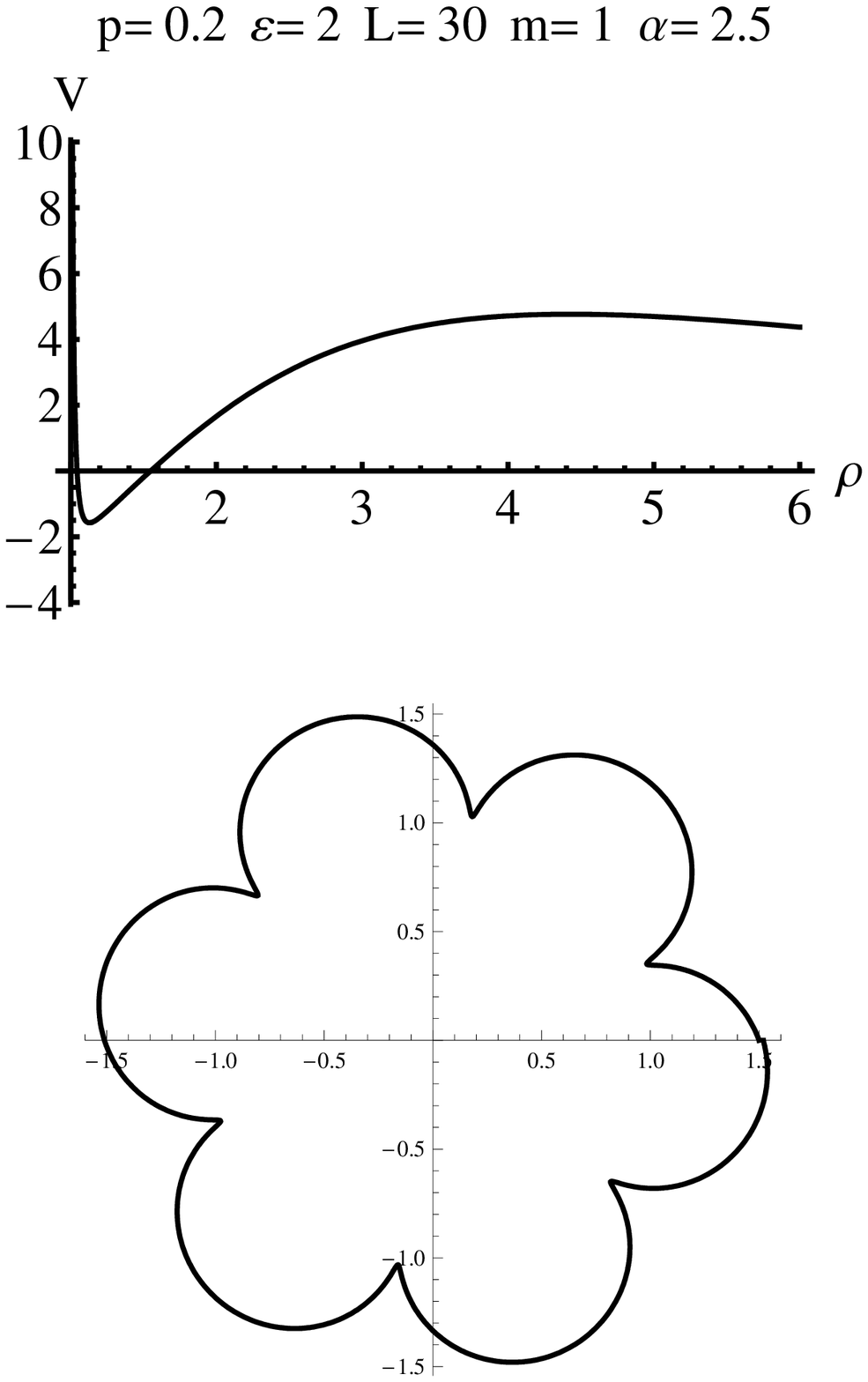}
\includegraphics[width=1.3in]{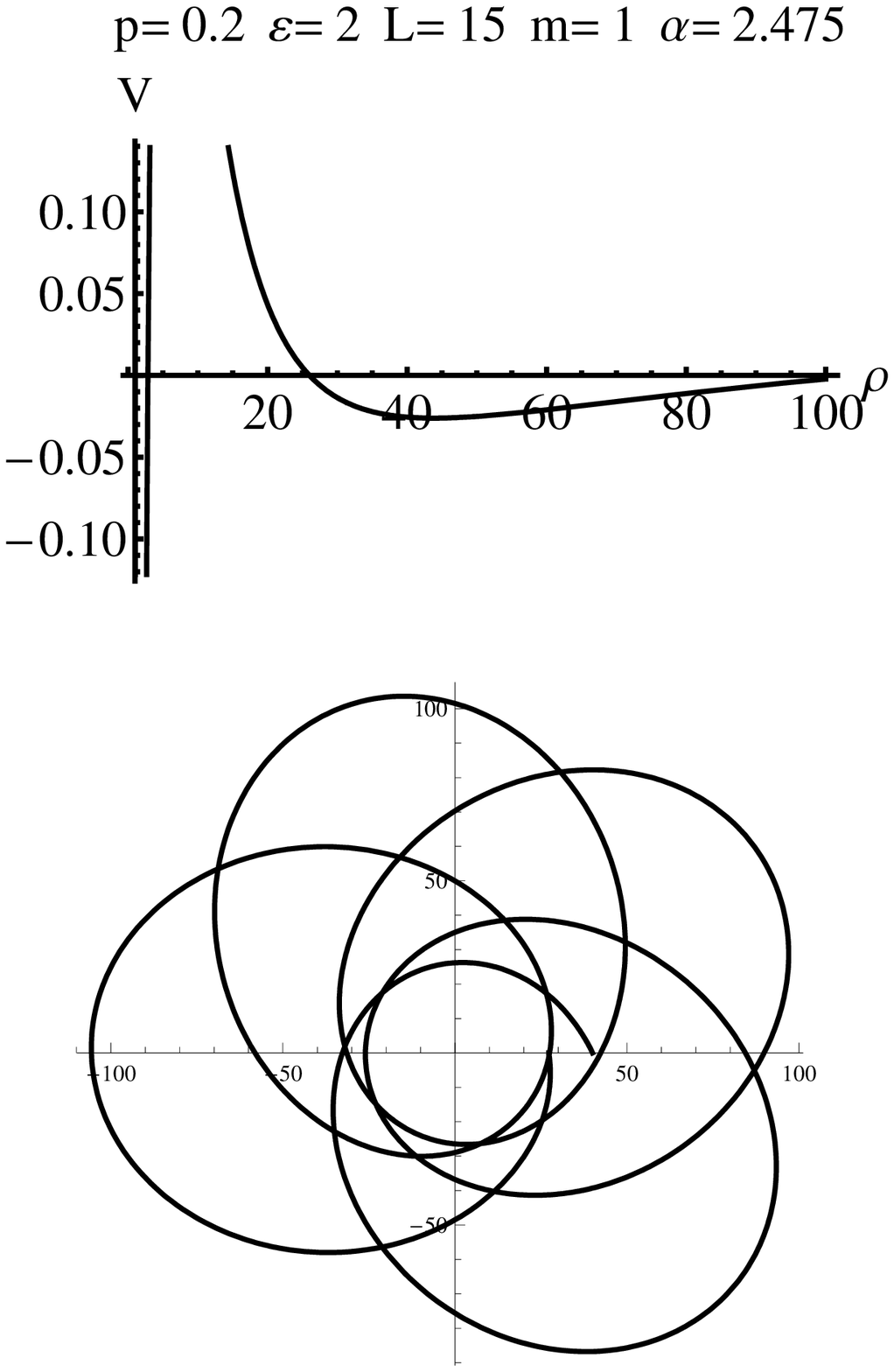}
\includegraphics[width=1.3in]{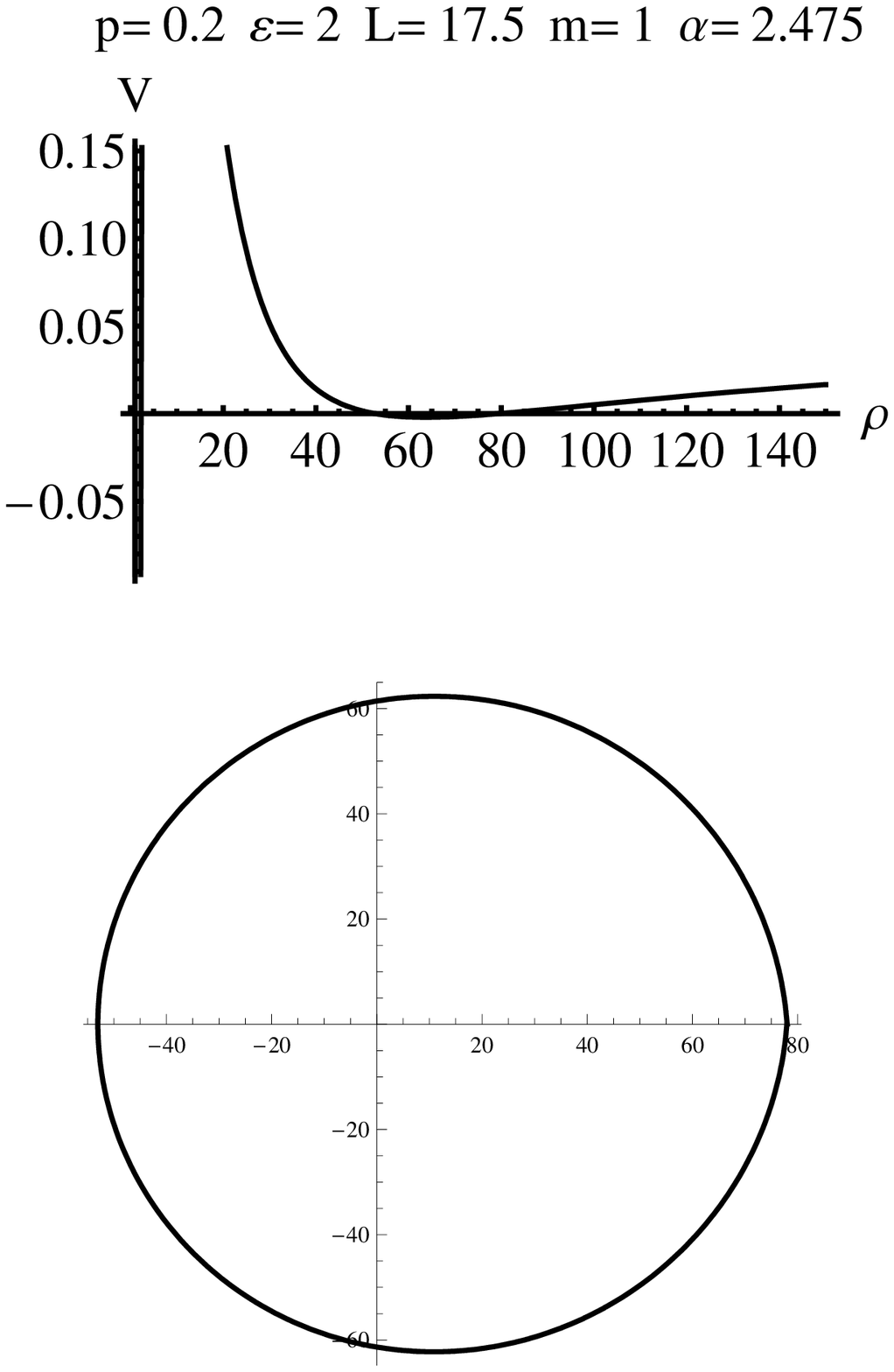}
\end{center}
\caption{\footnotesize{The effective potential and timelike
geodesic orbits in naked singularity spacetime with non-zero momentum
along the $z$-direction. The orbits are plotted in terms of
coordinates $x=\rho\cos\phi$ and $y=\rho\sin\phi$. If we take the geometry with non-zero momentum along the $z$-direction, each figure shows that the curves are projected on the $x$-$y$ plane.
The explicit values of momentum charge $\mathfrak{p}$ of the geometry and the energy $E$, angular momentum $L$, and momentum to energy ratio $\alpha$ of particle are given in the figures. }}
\label{fig:fig2}
\end{figure}

\begin{figure}[t]
\begin{center}
\includegraphics[width=1.3in]{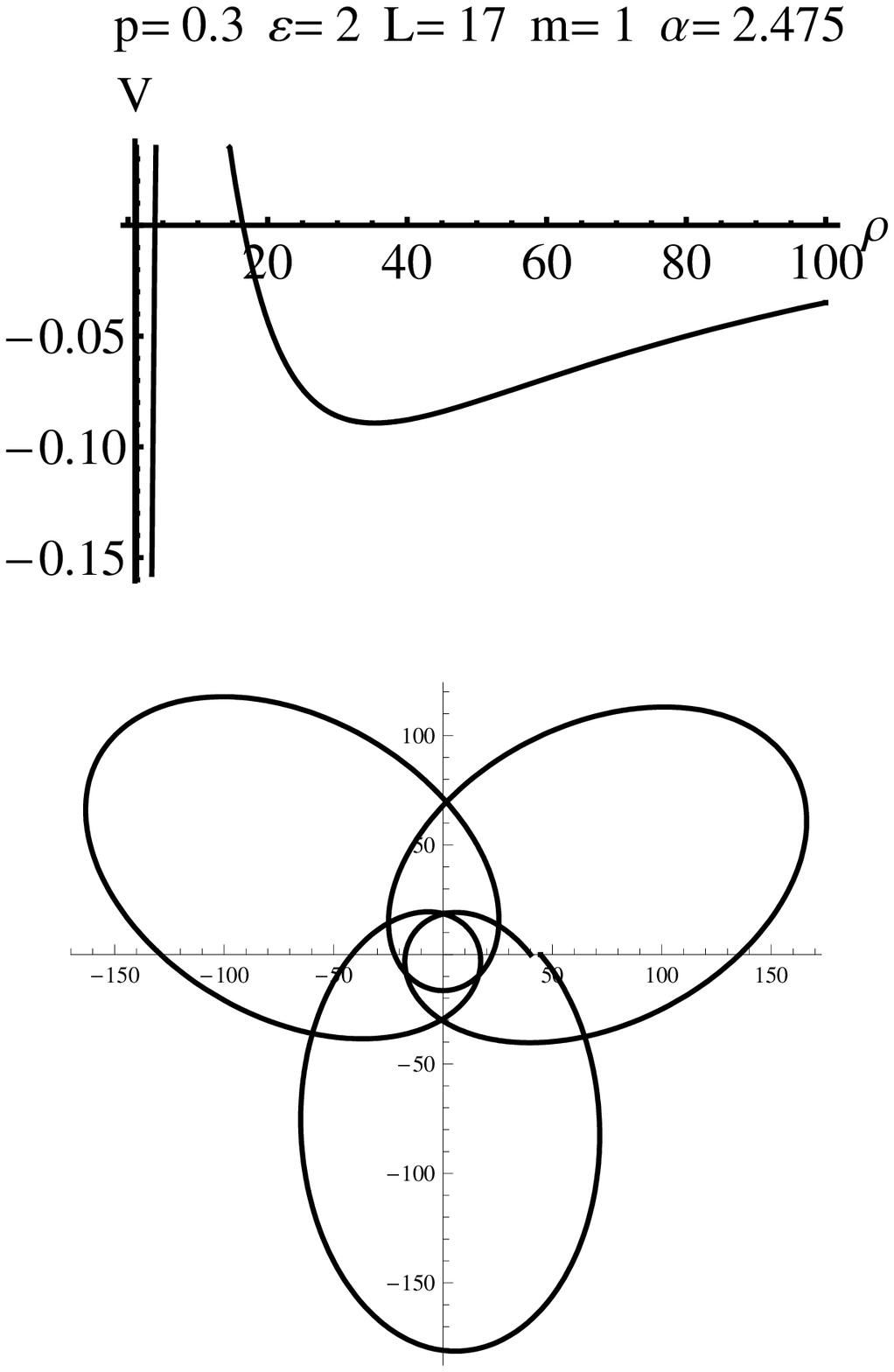}
\includegraphics[width=1.3in]{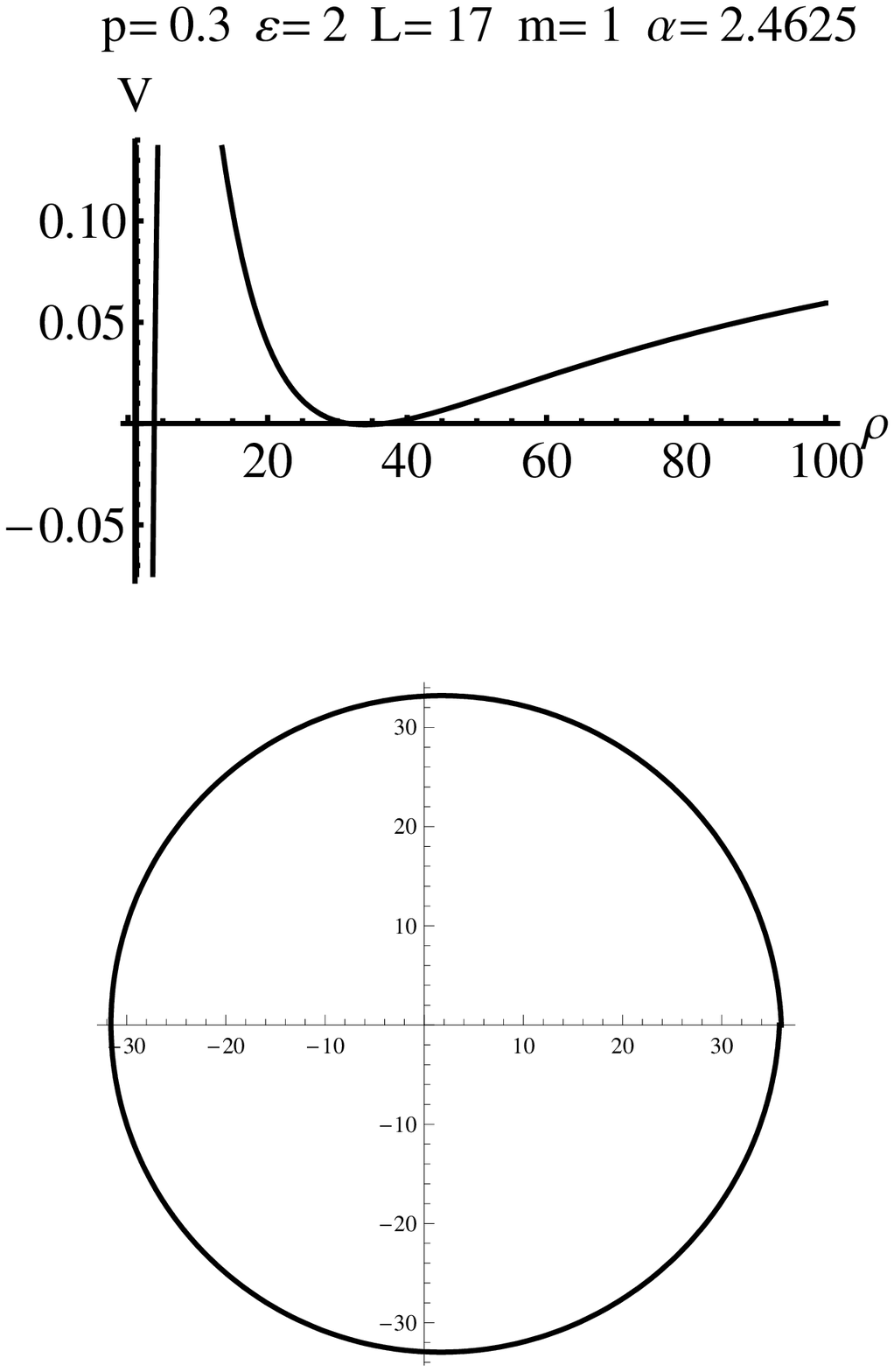}
\includegraphics[width=1.3in]{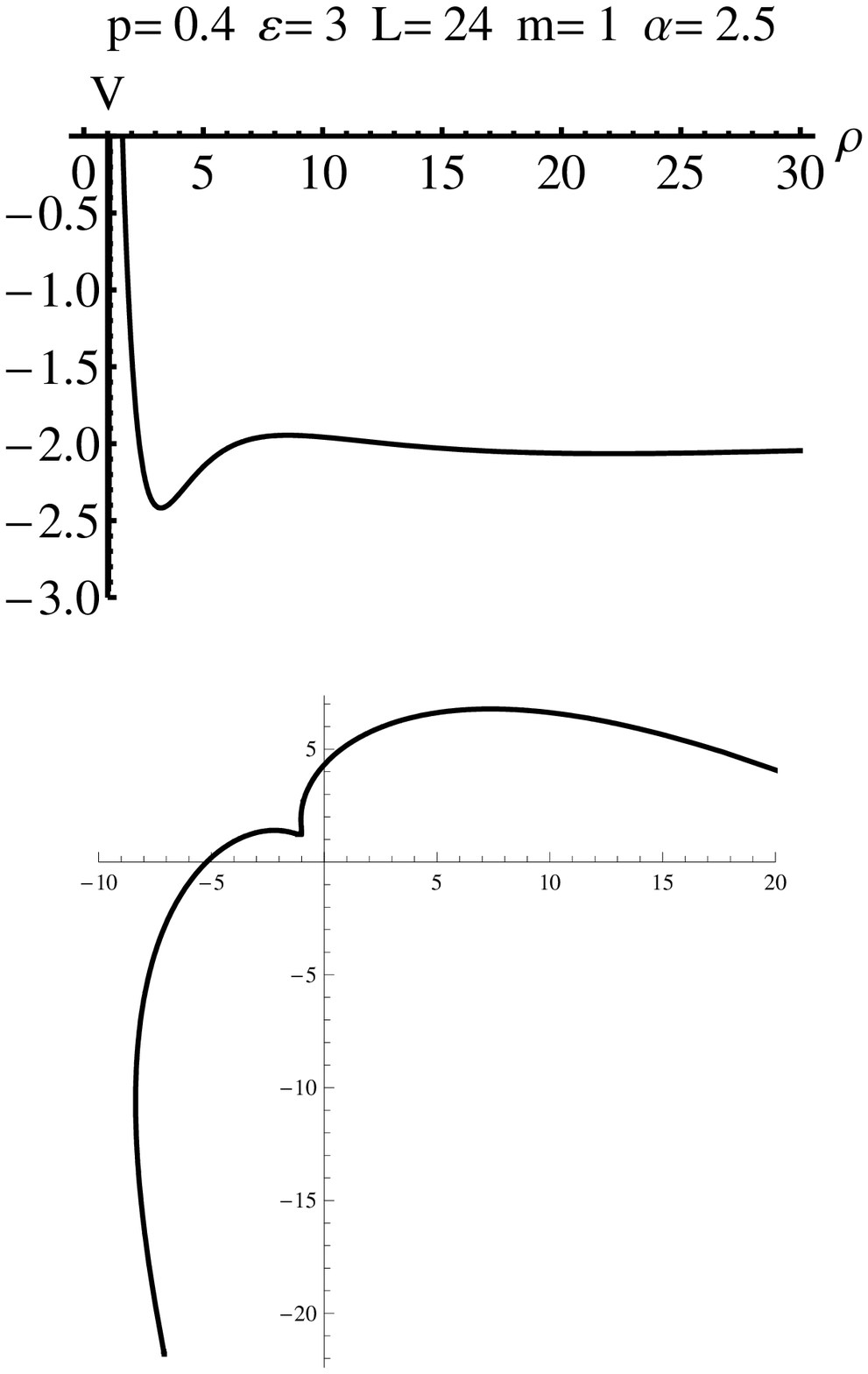}
\includegraphics[width=1.3in]{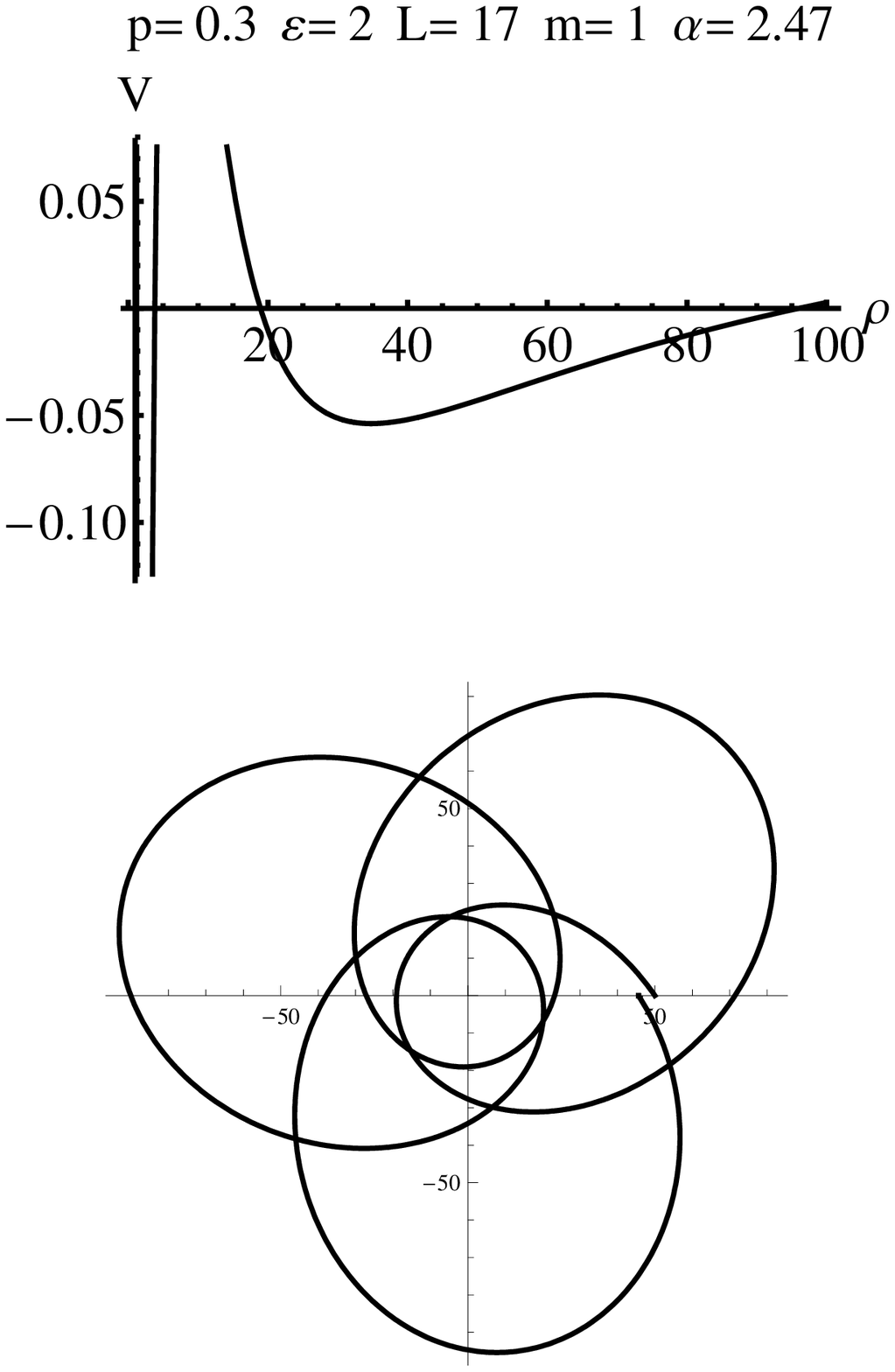}
\end{center}
\caption{\footnotesize{The effective potential and timelike
geodesic orbits in black string spacetime with non-zero momentum along
the $z$-direction.  }} \label{fig:fig3}
\end{figure}

\begin{figure}[h]
\begin{center}
\includegraphics[width=1.3in]{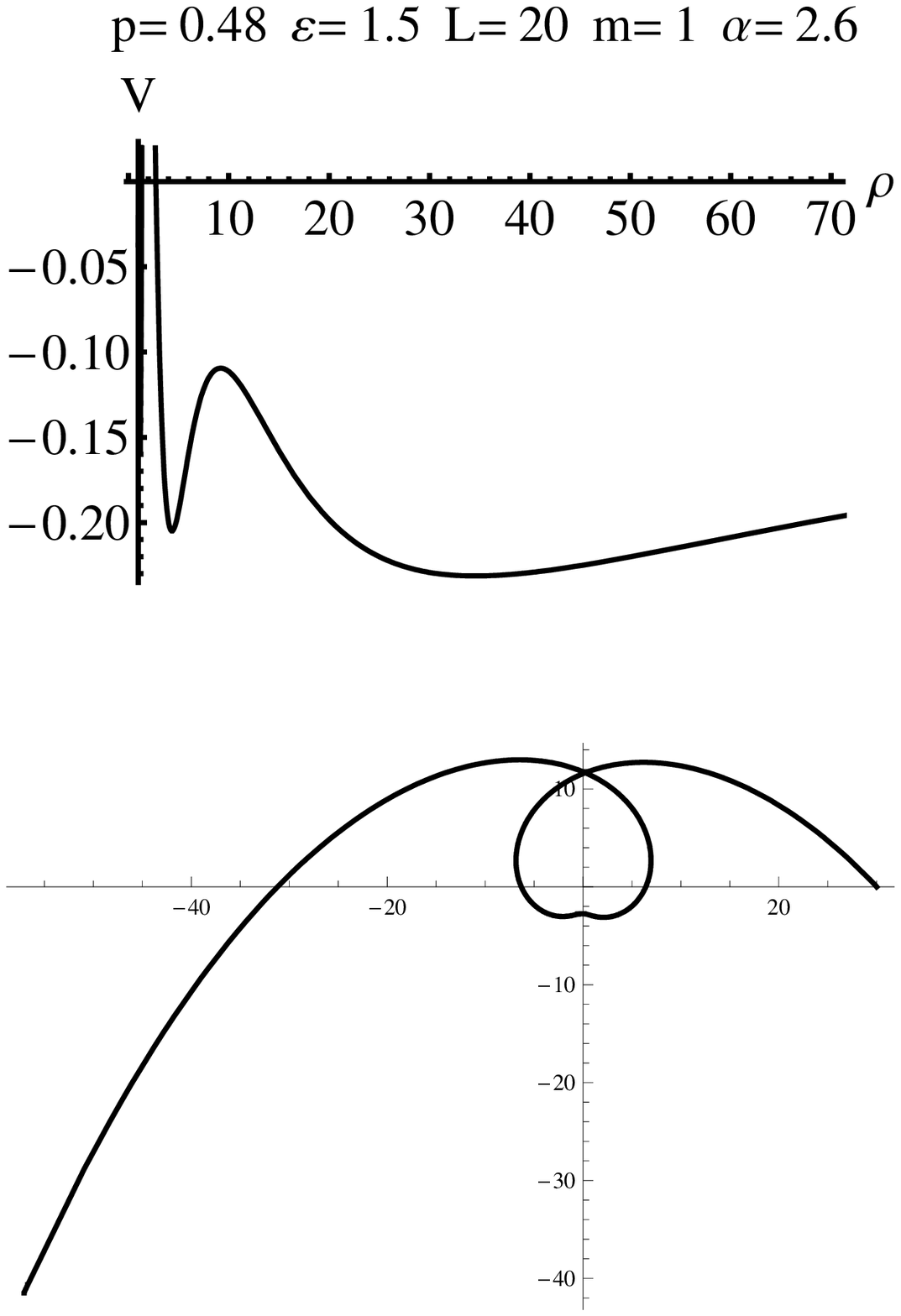}
\includegraphics[width=1.3in]{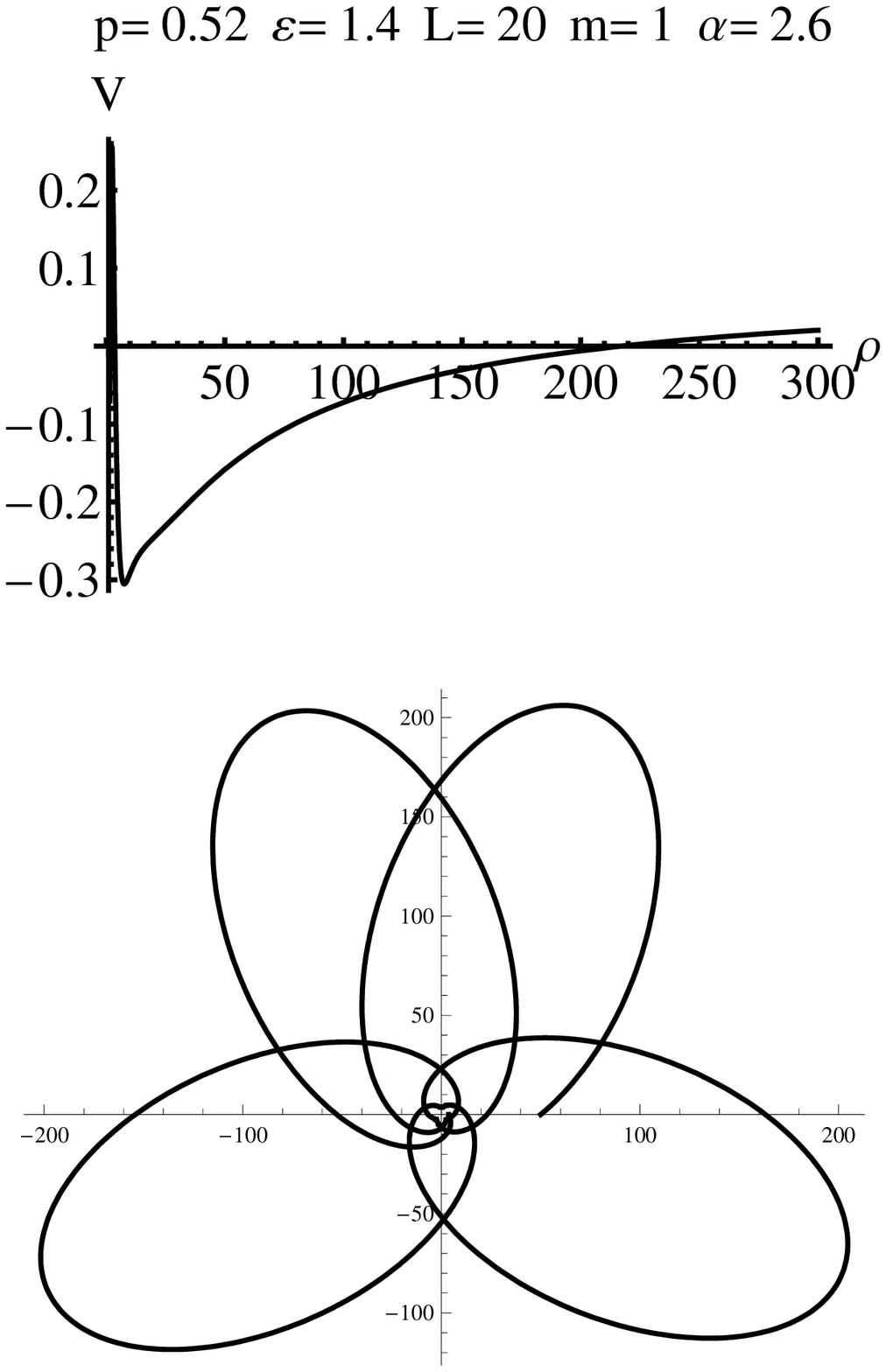}
\includegraphics[width=1.3in]{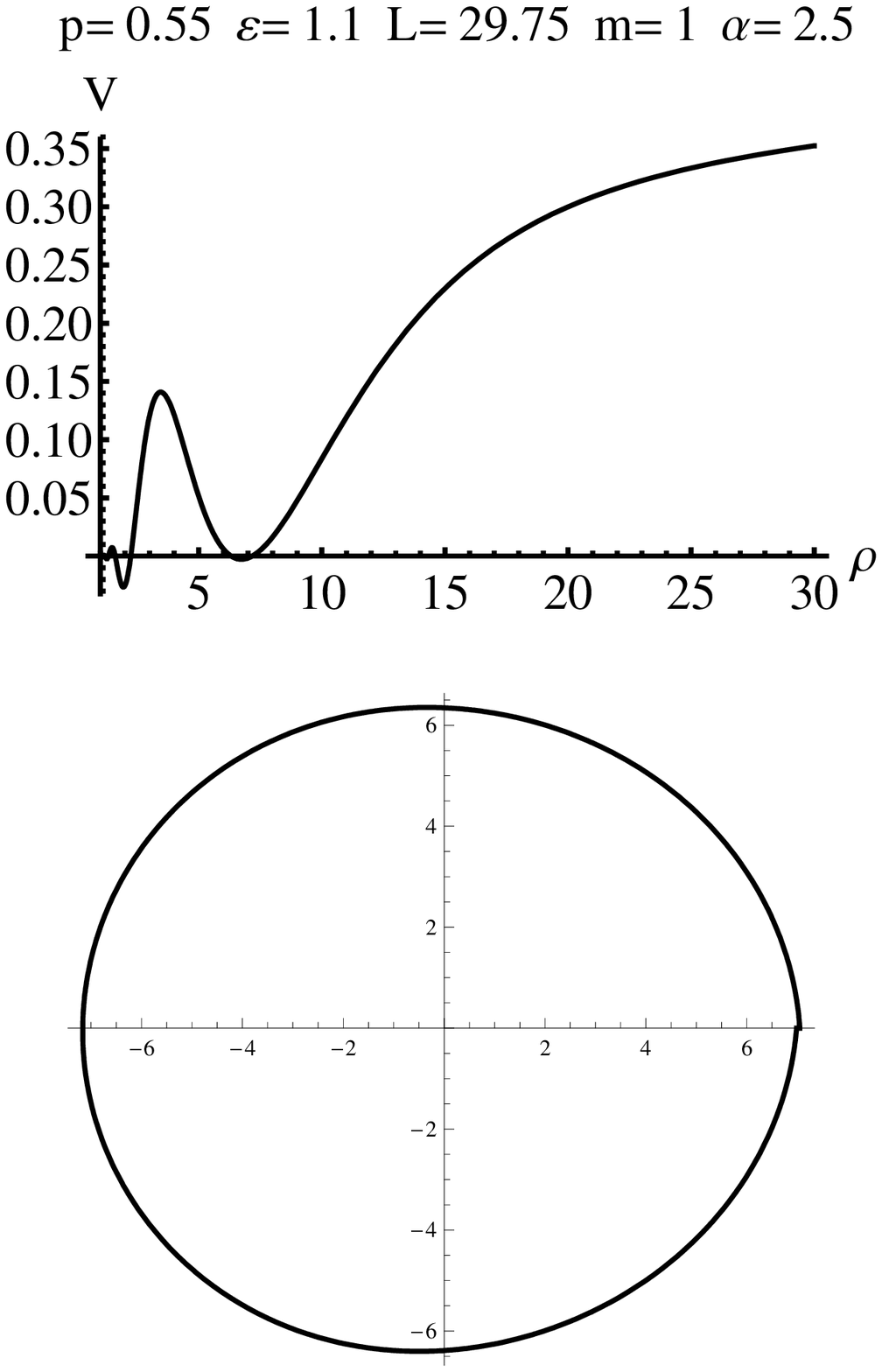}
\includegraphics[width=1.3in]{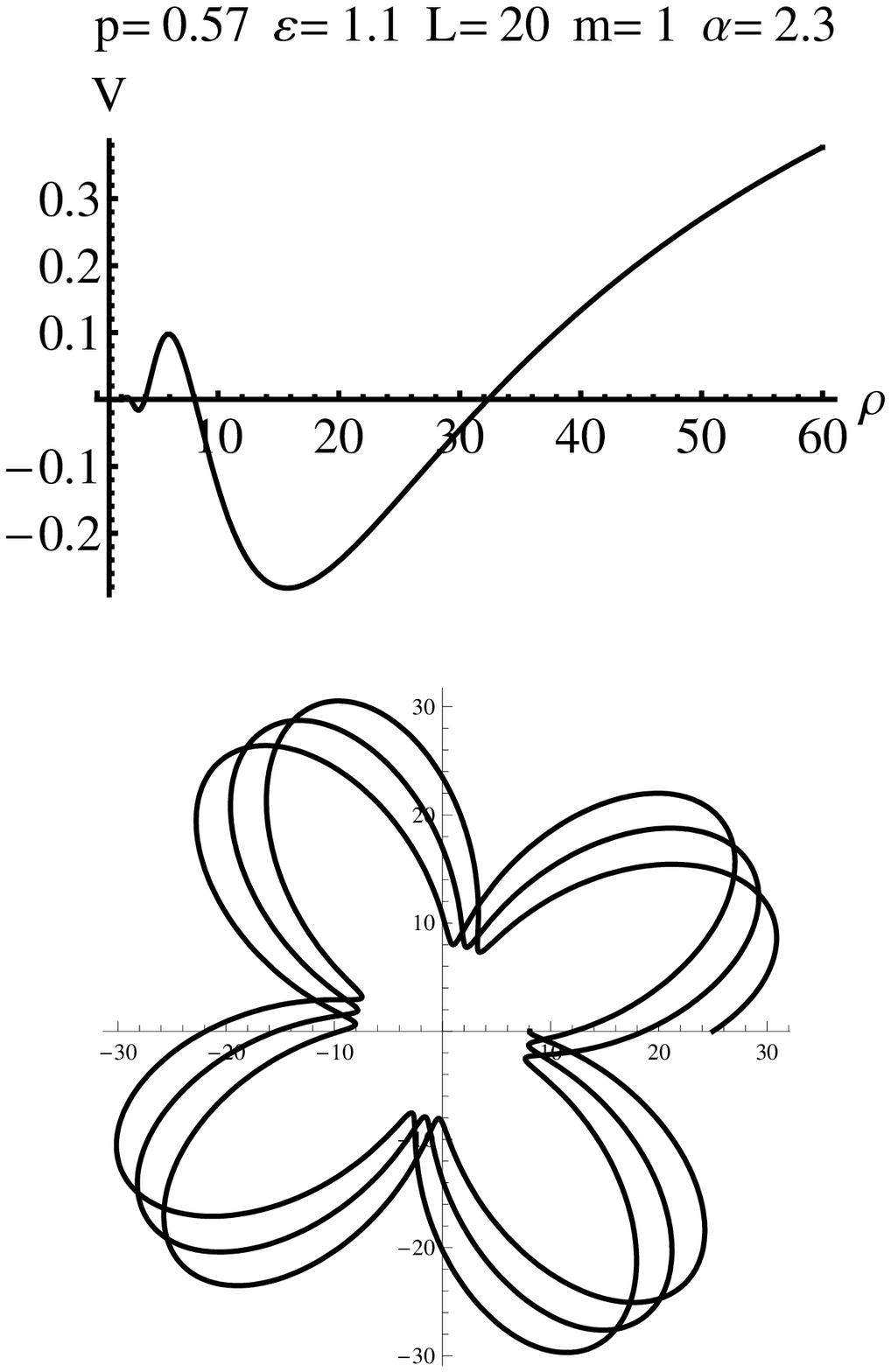}
\end{center}
\caption{\footnotesize{The effective potential and timelike
geodesic orbits in wormhole geometry with non-zero momentum
along the $z$-direction.  }} \label{fig:fig4}
\end{figure}

\begin{figure}
\begin{center}
\includegraphics[width=1.3in]{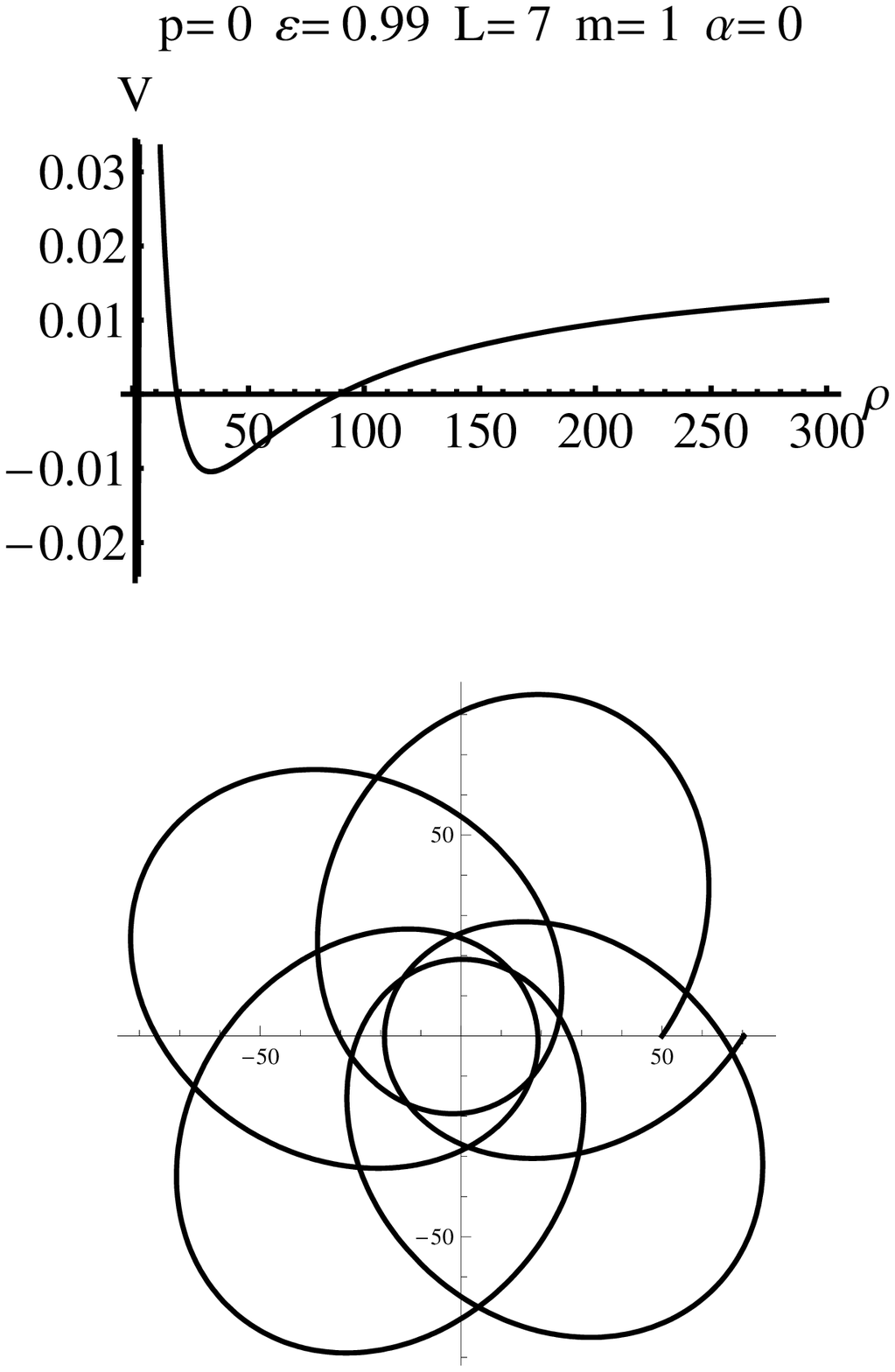}
\includegraphics[width=1.3in]{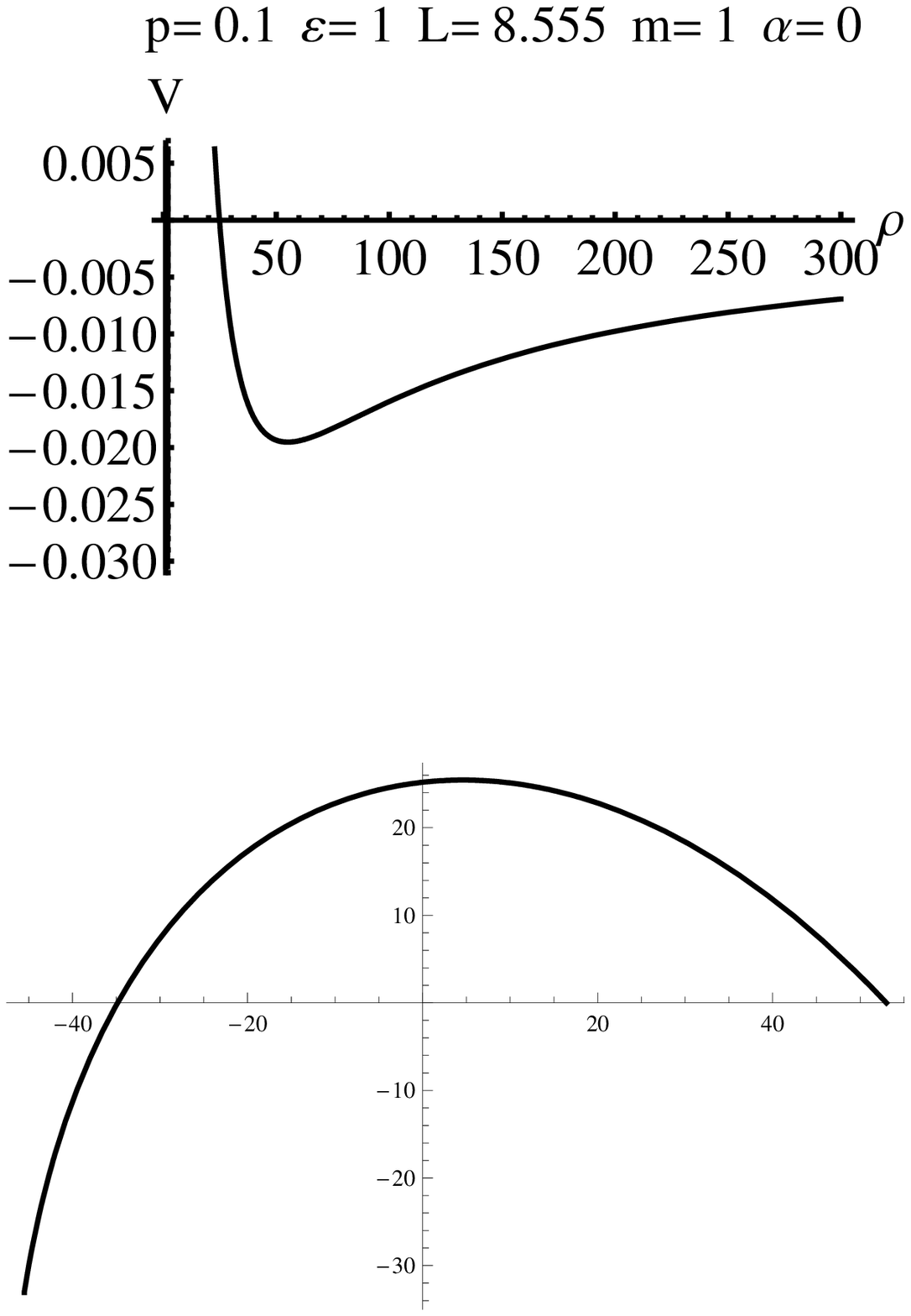}
\includegraphics[width=1.3in]{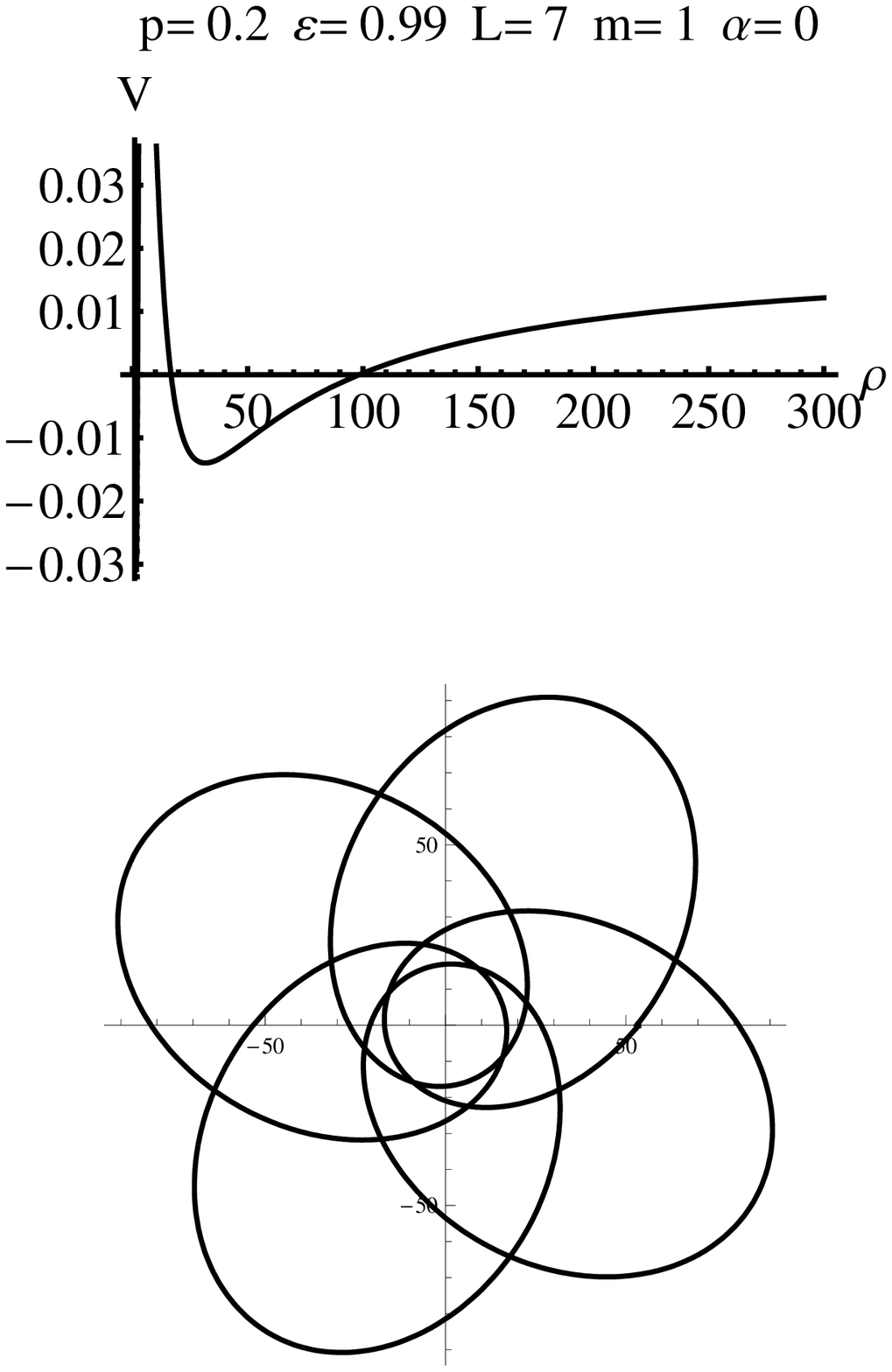}
\includegraphics[width=1.3in]{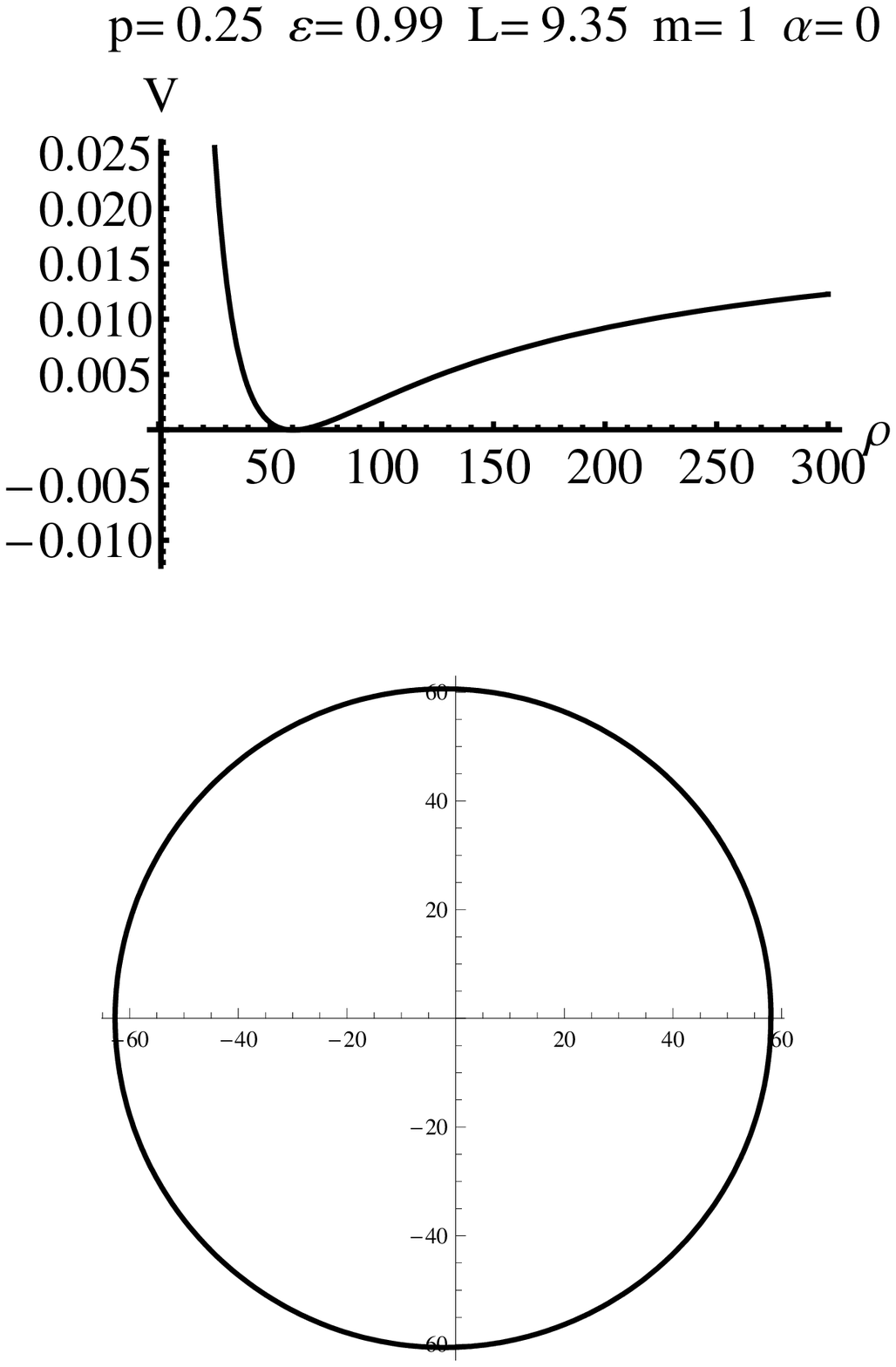}
\end{center}
\caption{\footnotesize{The effective potential and timelike
geodesic orbits in naked singular spacetime with zero momentum along
the $z$-direction. }} \label{fig:fig5}
\end{figure}

\begin{figure}[h]
\begin{center}
\includegraphics[width=1.3in]{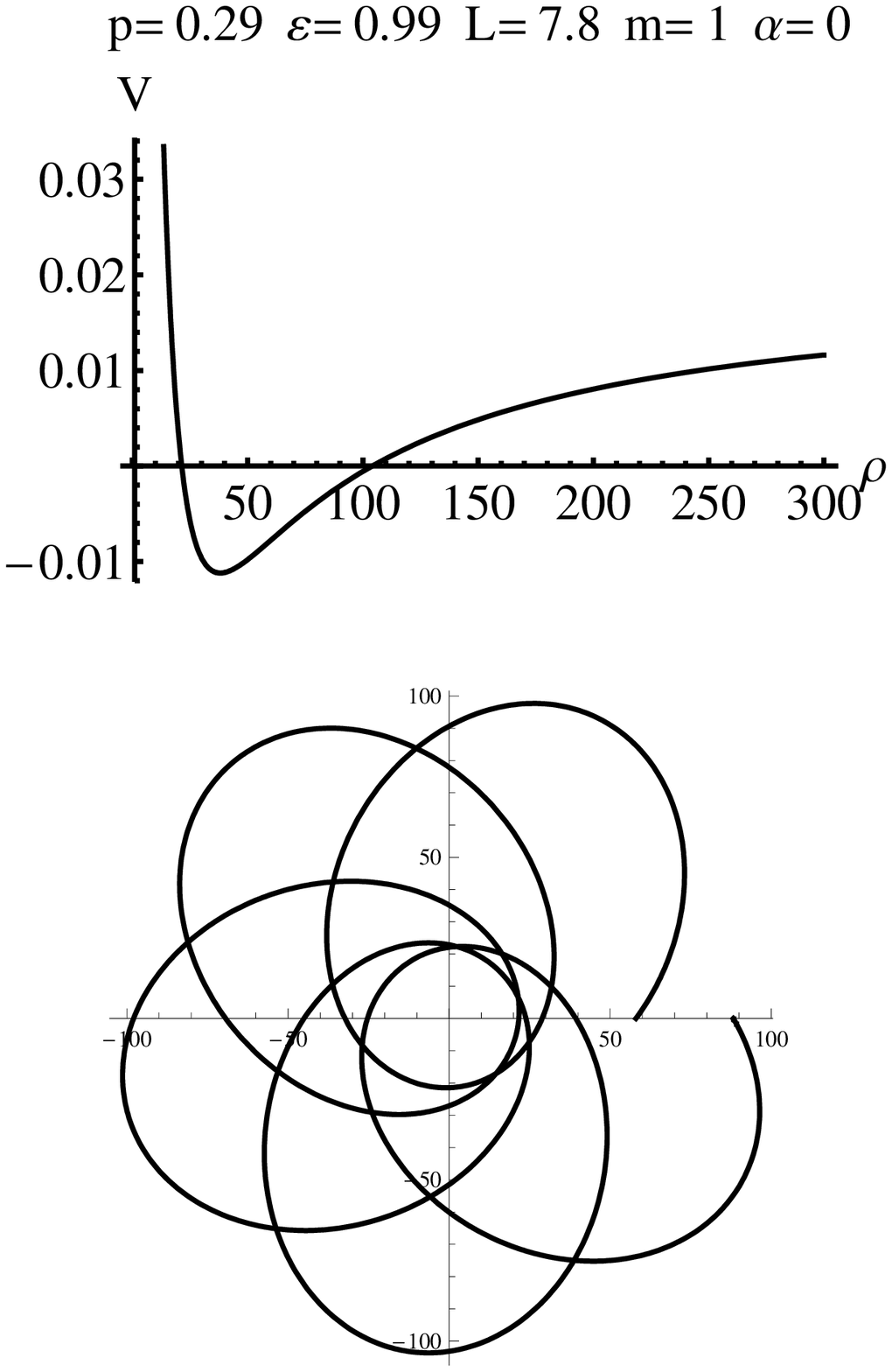}
\includegraphics[width=1.3in]{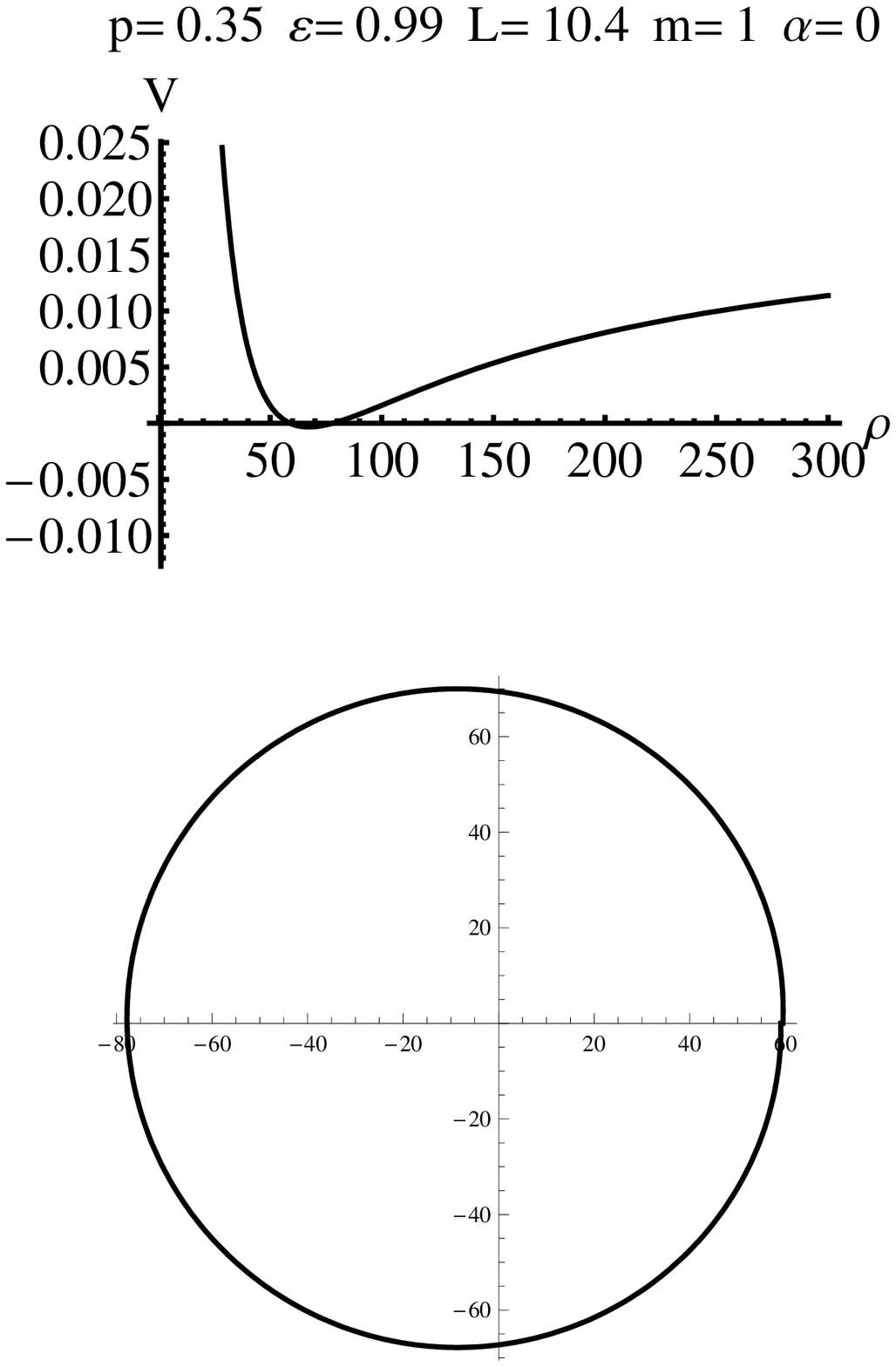}
\includegraphics[width=1.3in]{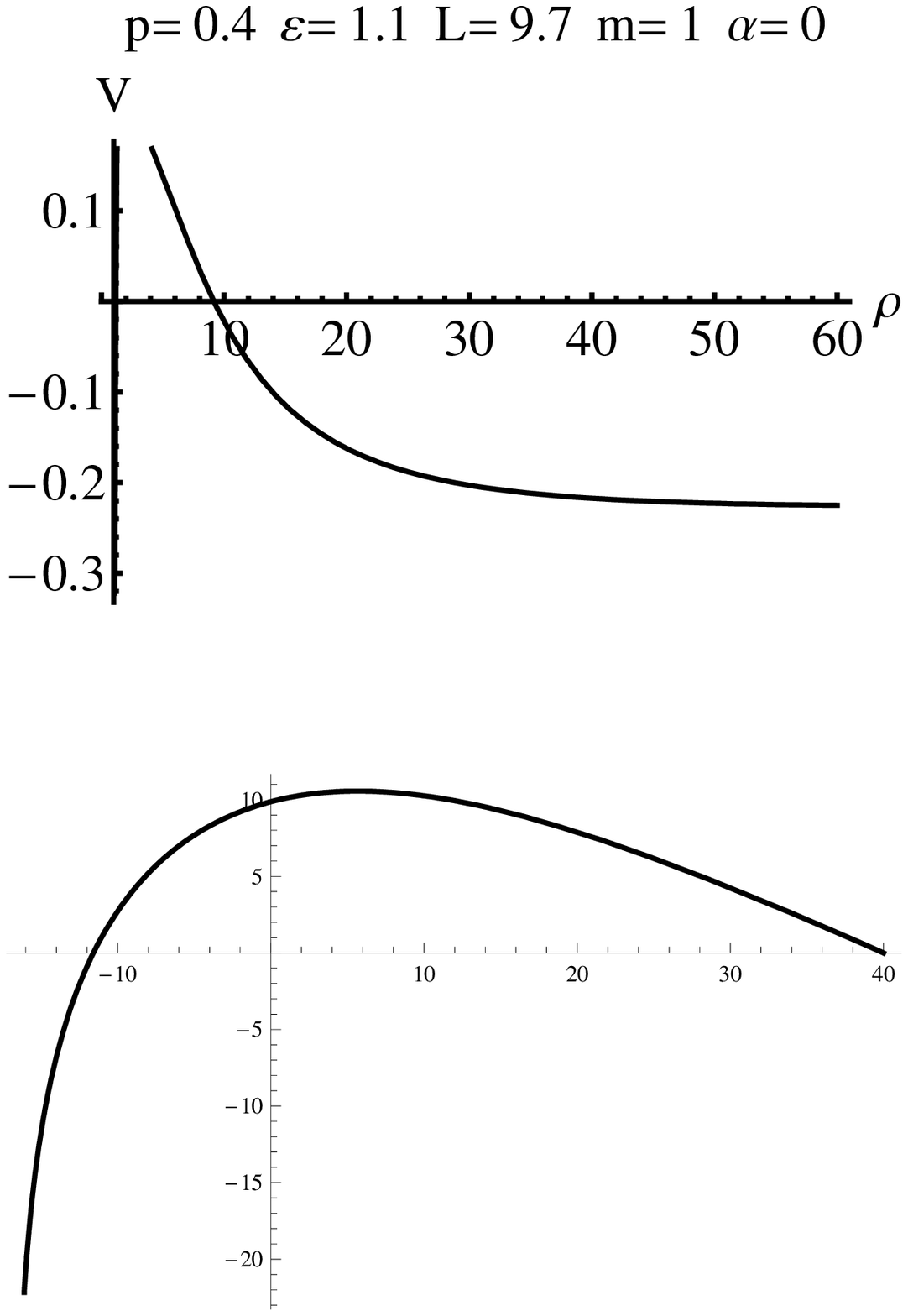}
\includegraphics[width=1.3in]{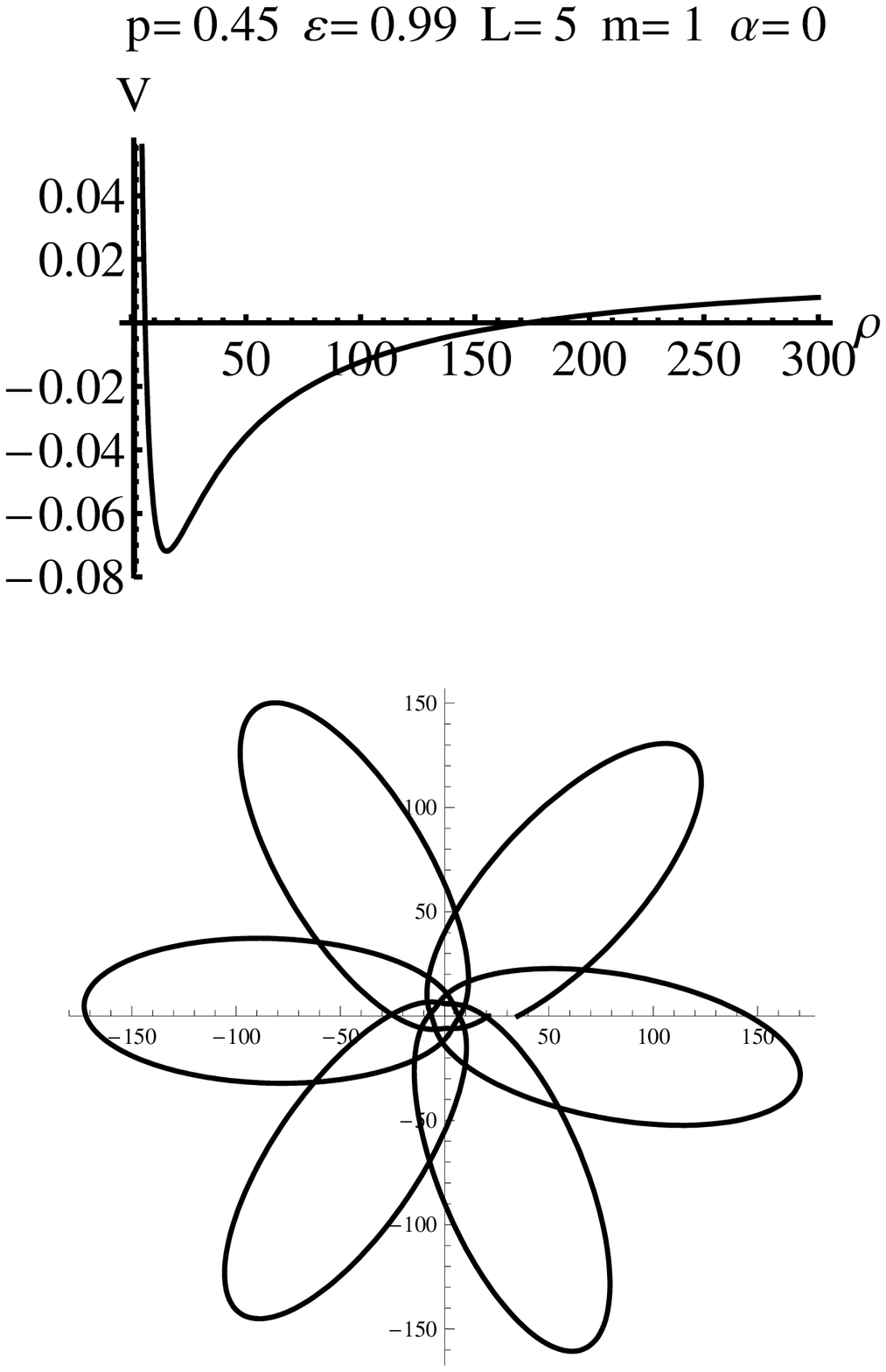}
\end{center}
\caption{\footnotesize{The effective potential and timelike
geodesic orbits black string spacetime with zero momentum along the
$z$-direction. }} \label{fig:fig6}
\end{figure}

\begin{figure}[h]
\begin{center}
\includegraphics[width=1.3in]{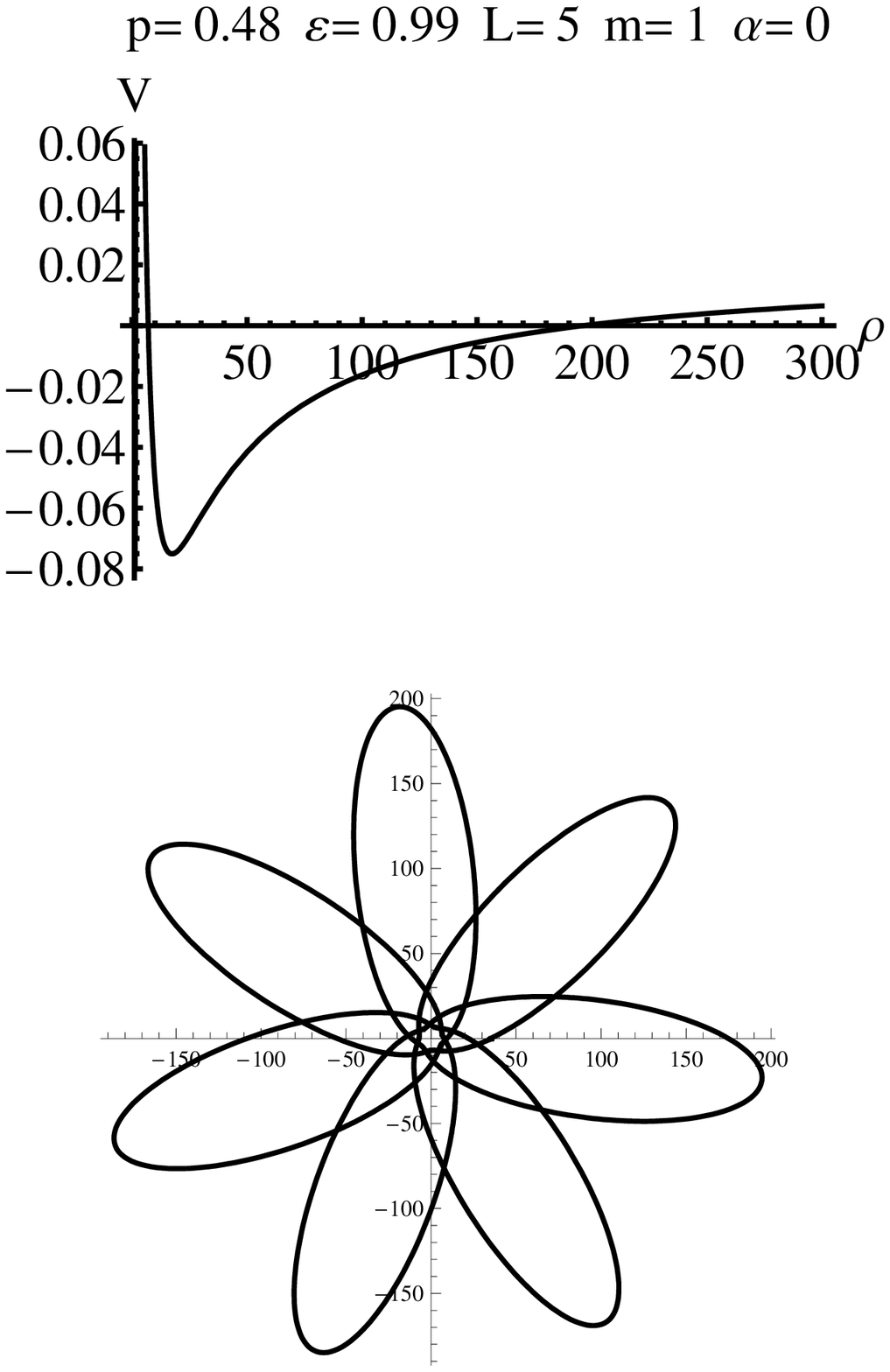}
\includegraphics[width=1.3in]{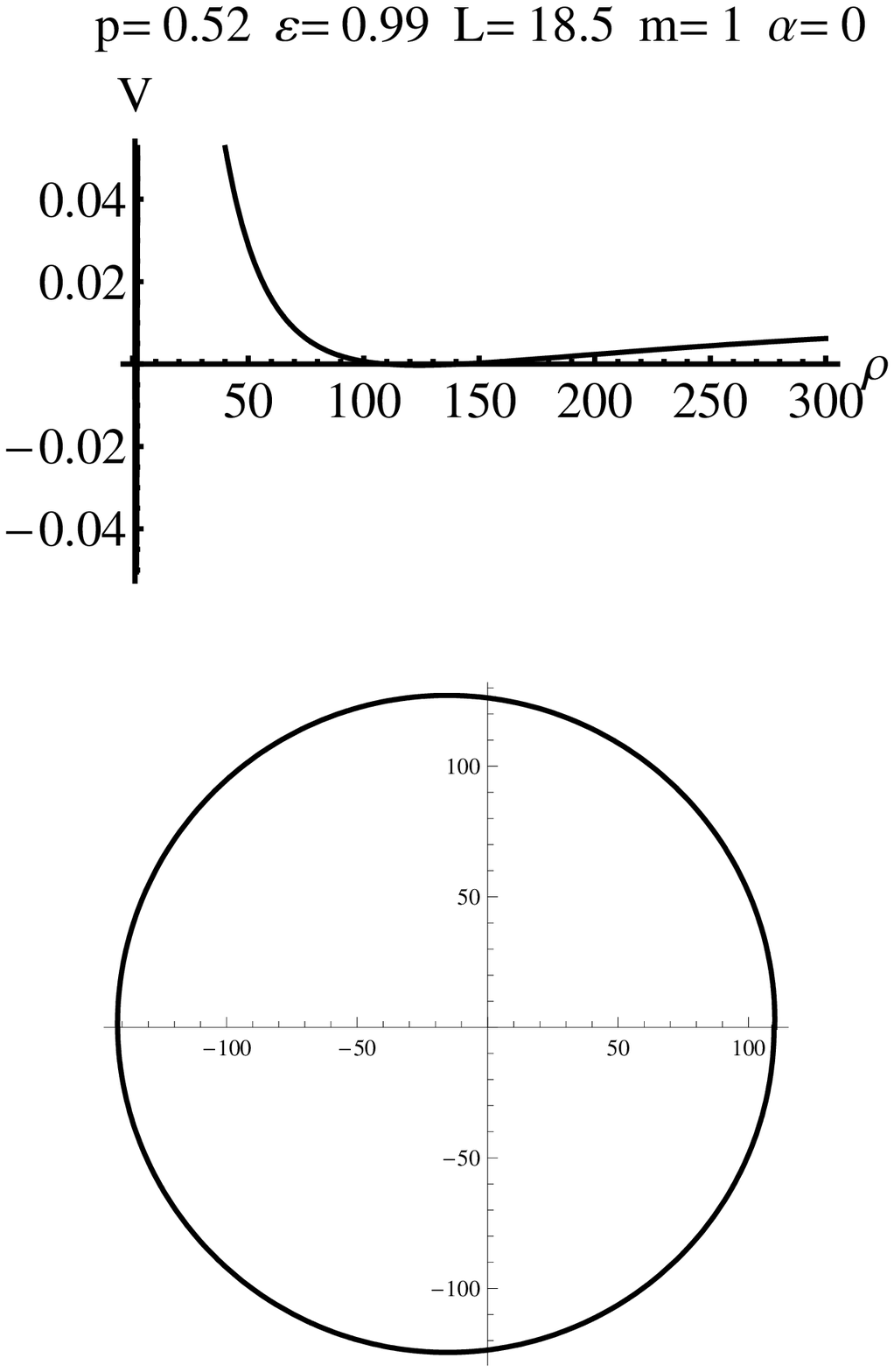}
\includegraphics[width=1.3in]{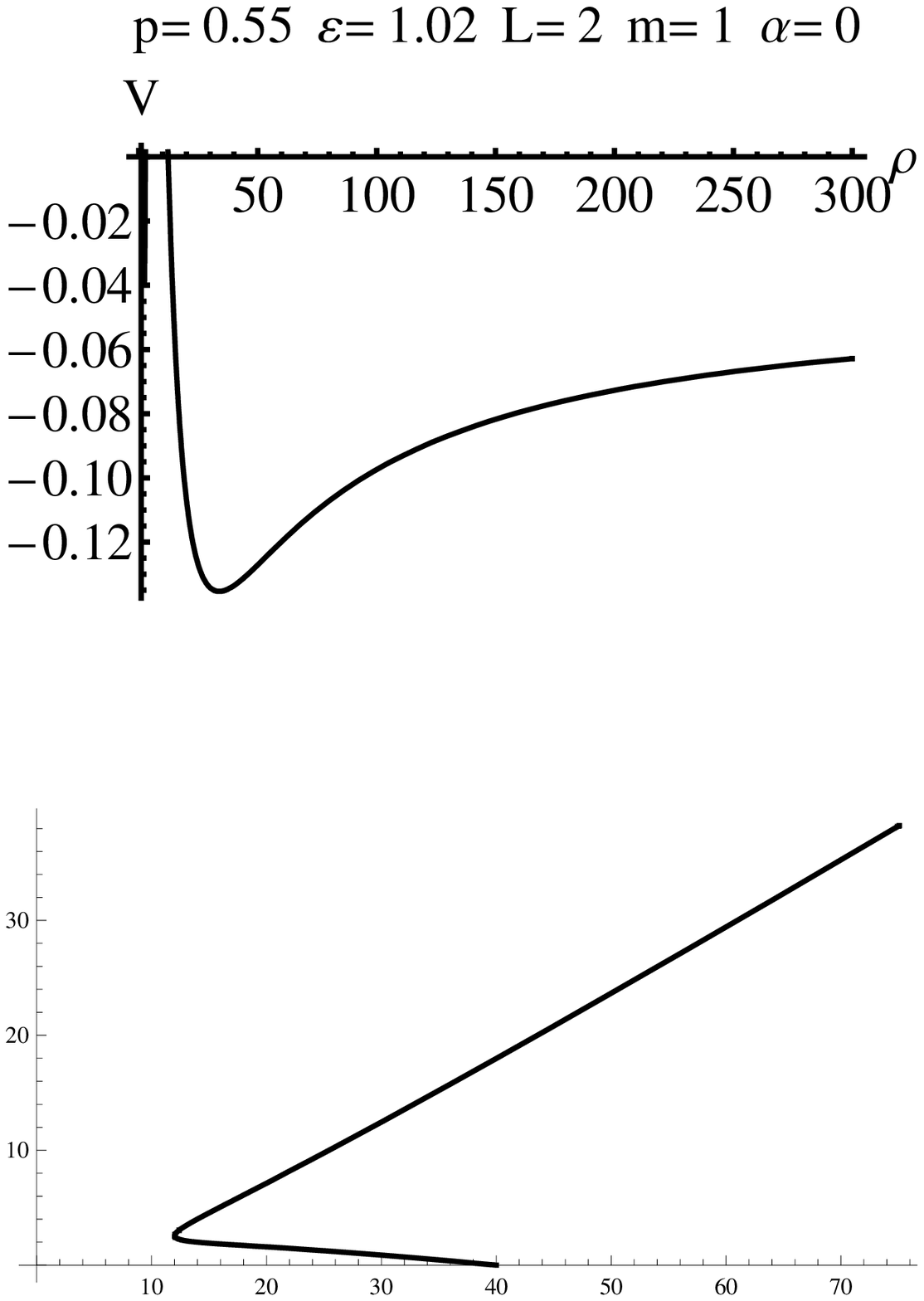}
\includegraphics[width=1.3in]{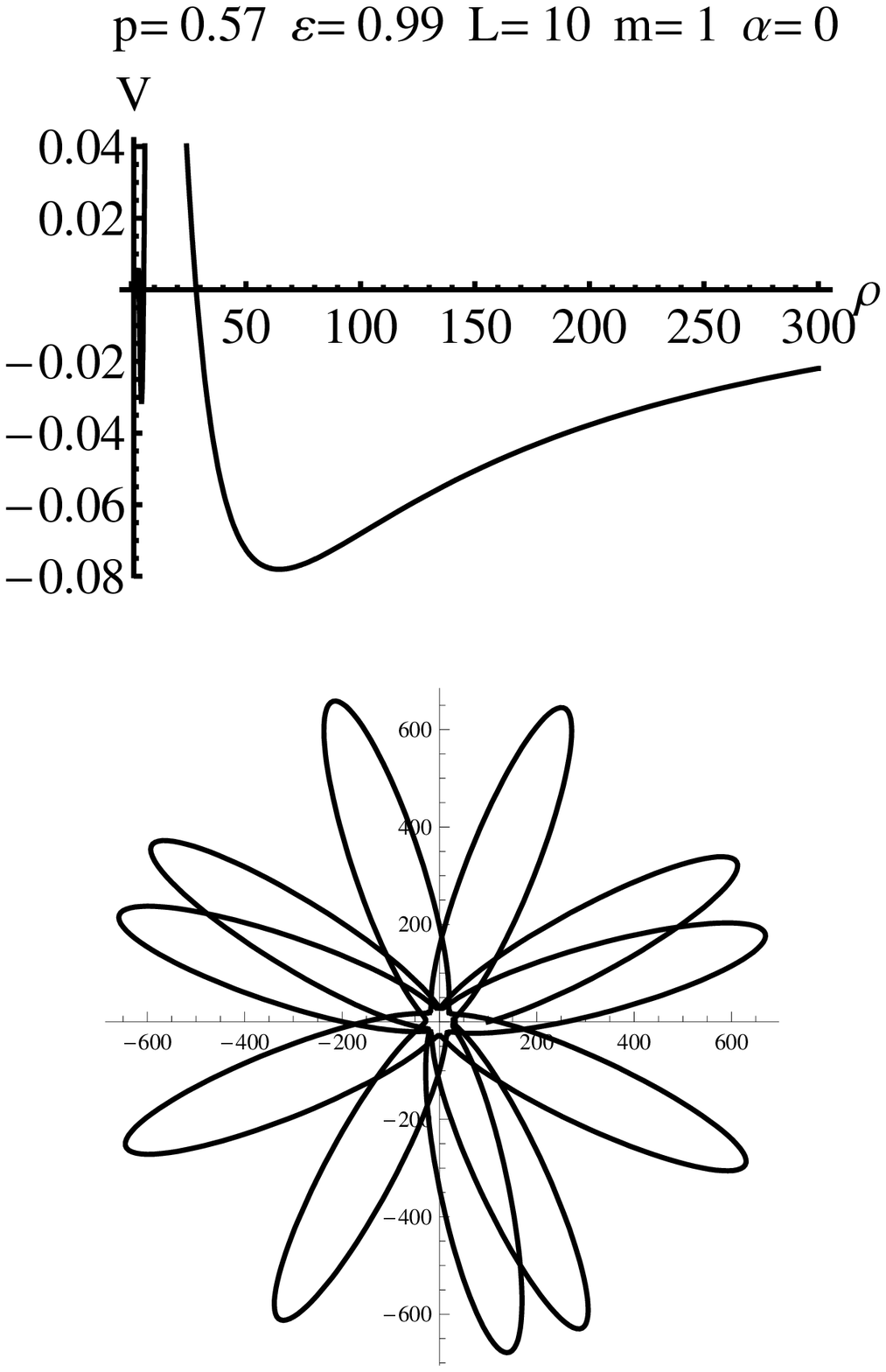}
\end{center}
\caption{\footnotesize{The effective potential and timelike
geodesic orbits in wormhole geometry with zero momentum along
the $z$-direction. }} \label{fig:fig7}
\end{figure}

\begin{figure}
\begin{center}
\includegraphics[width=1.3in]{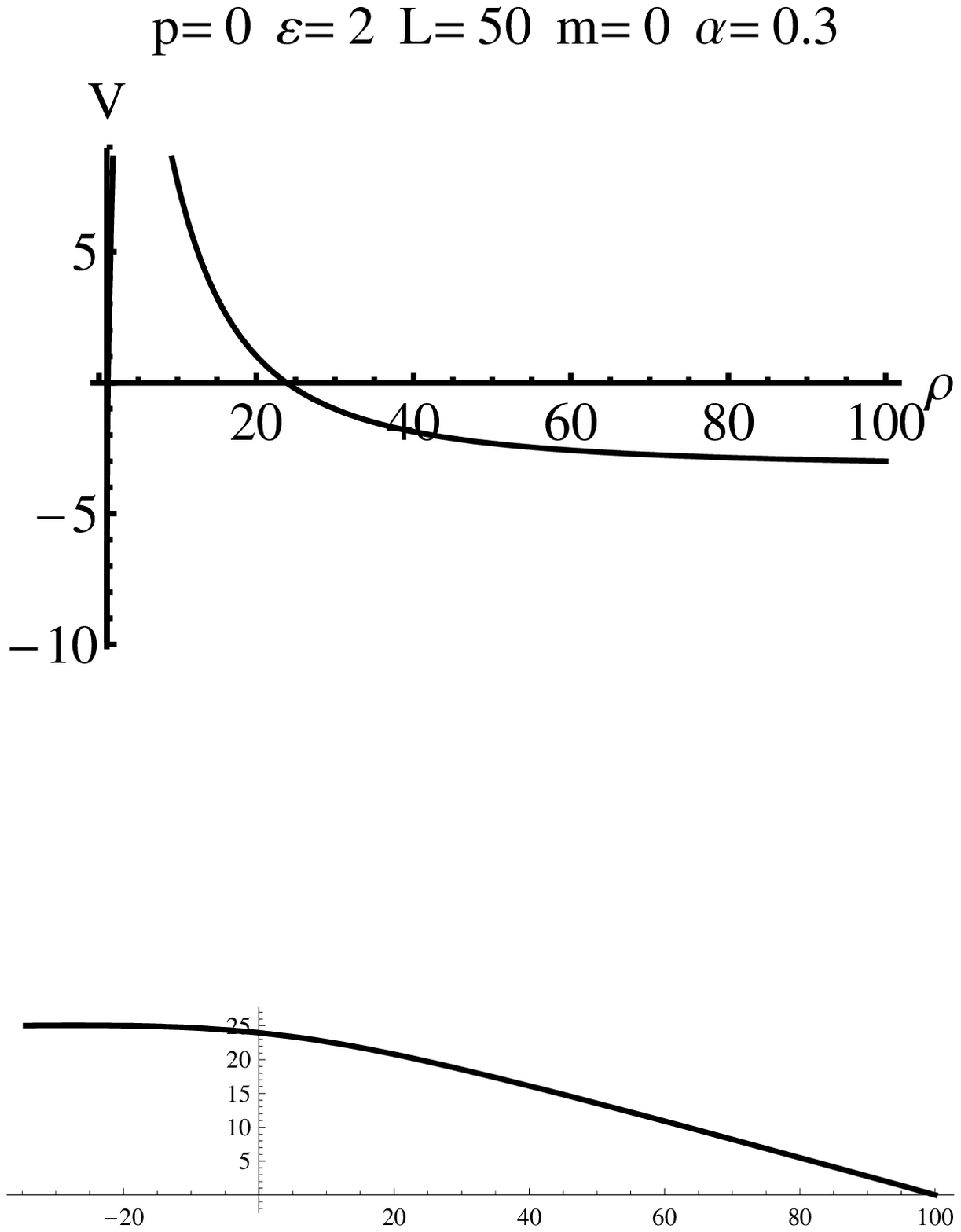}
\includegraphics[width=1.3in]{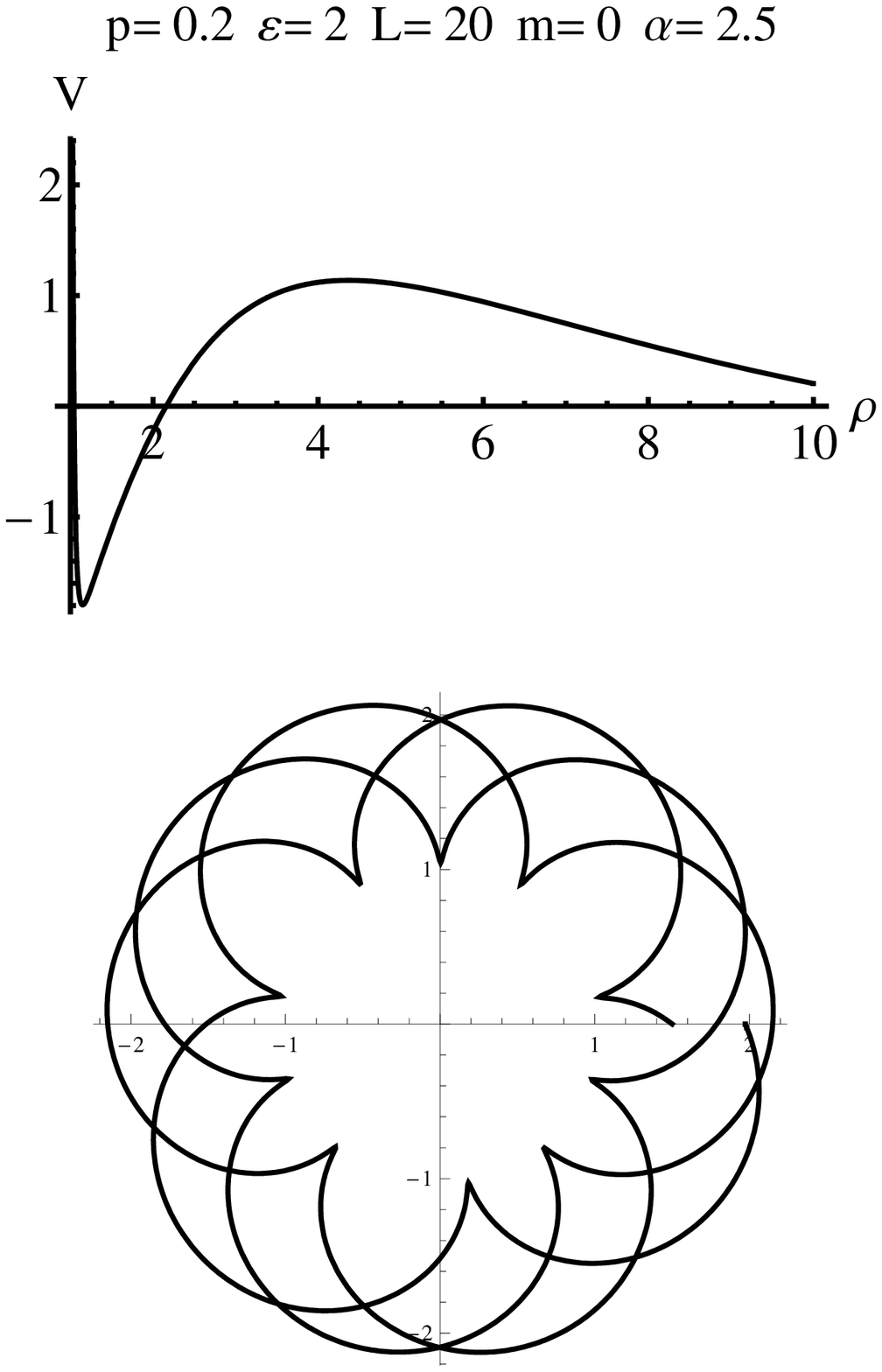}
\includegraphics[width=1.3in]{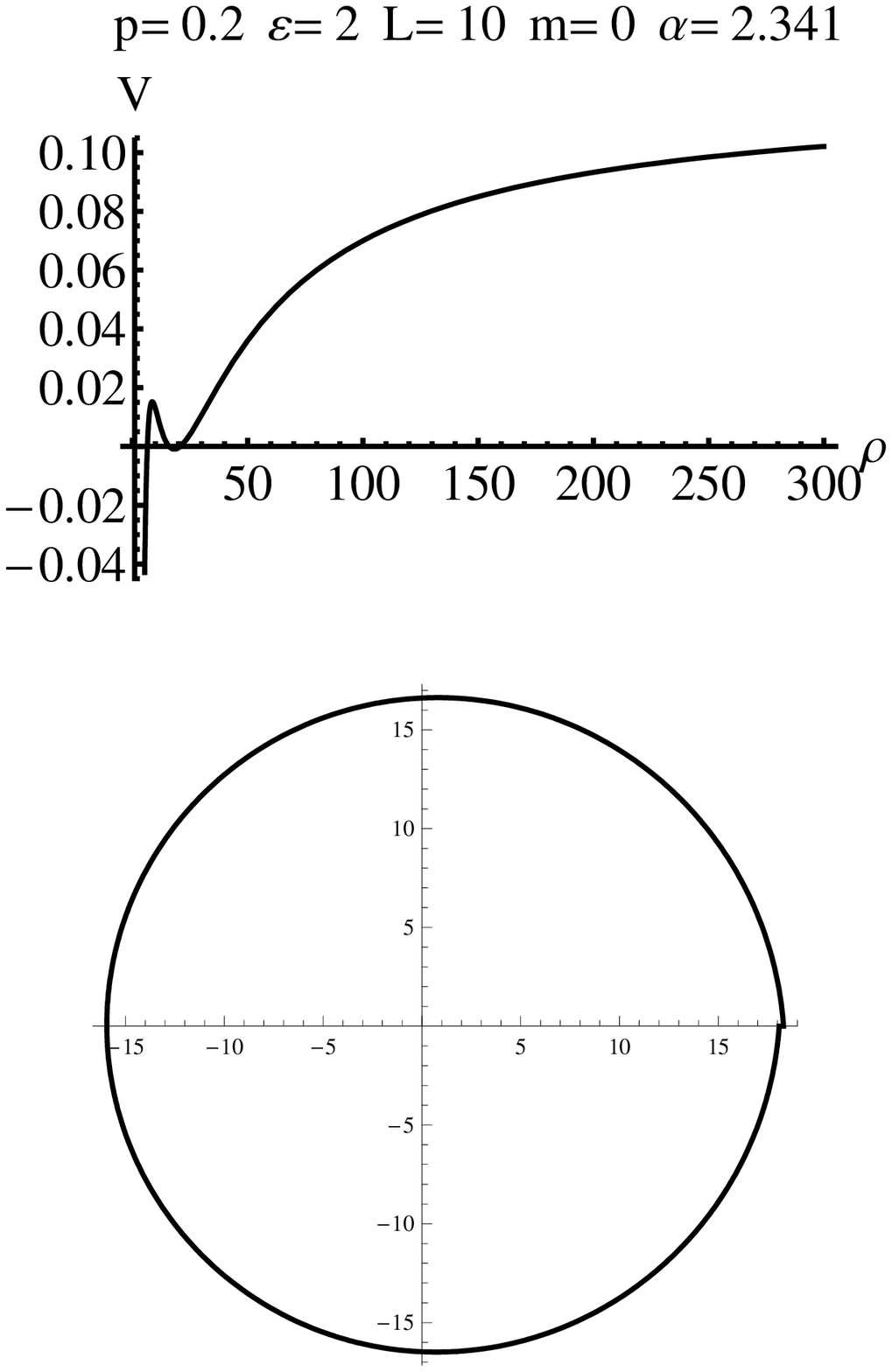}
\includegraphics[width=1.3in]{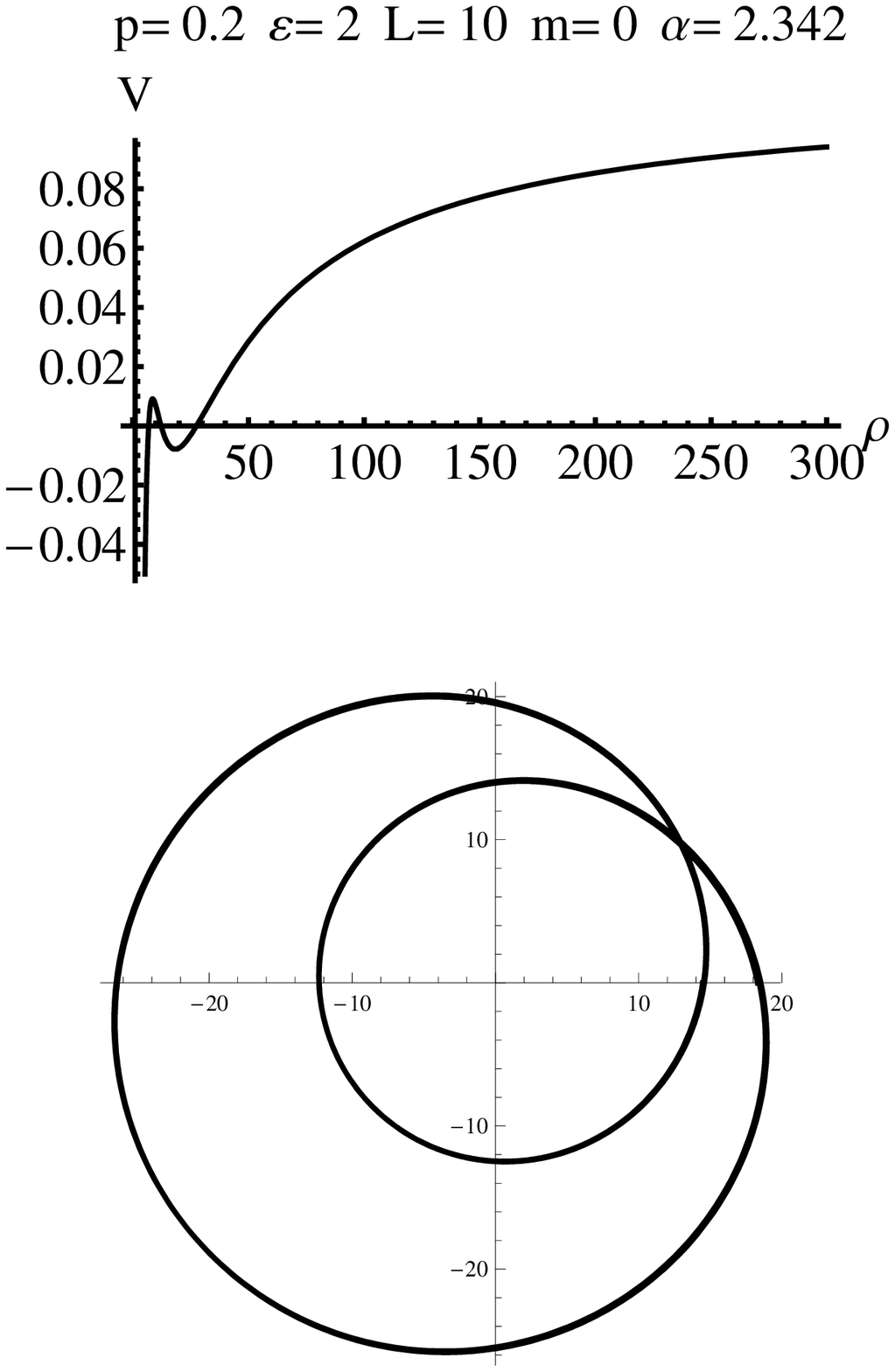}
\end{center}
\caption{\footnotesize{The effective potential and lightlike
geodesic orbits in naked singular spacetime with non-zero momentum along
the $z$-direction. }} \label{fig:fig8}
\end{figure}

\begin{figure}
\begin{center}
\includegraphics[width=1.3in]{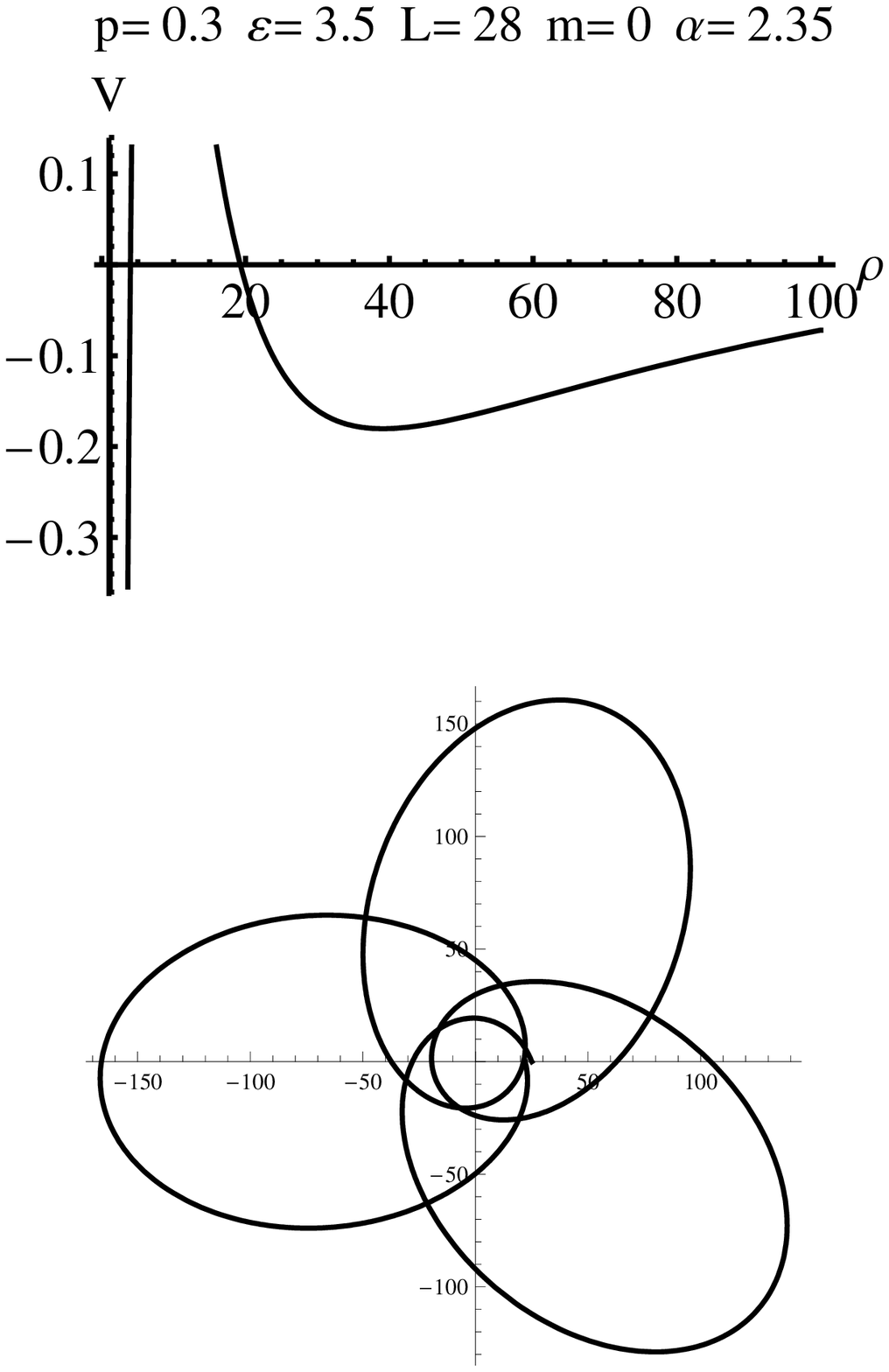}
\includegraphics[width=1.3in]{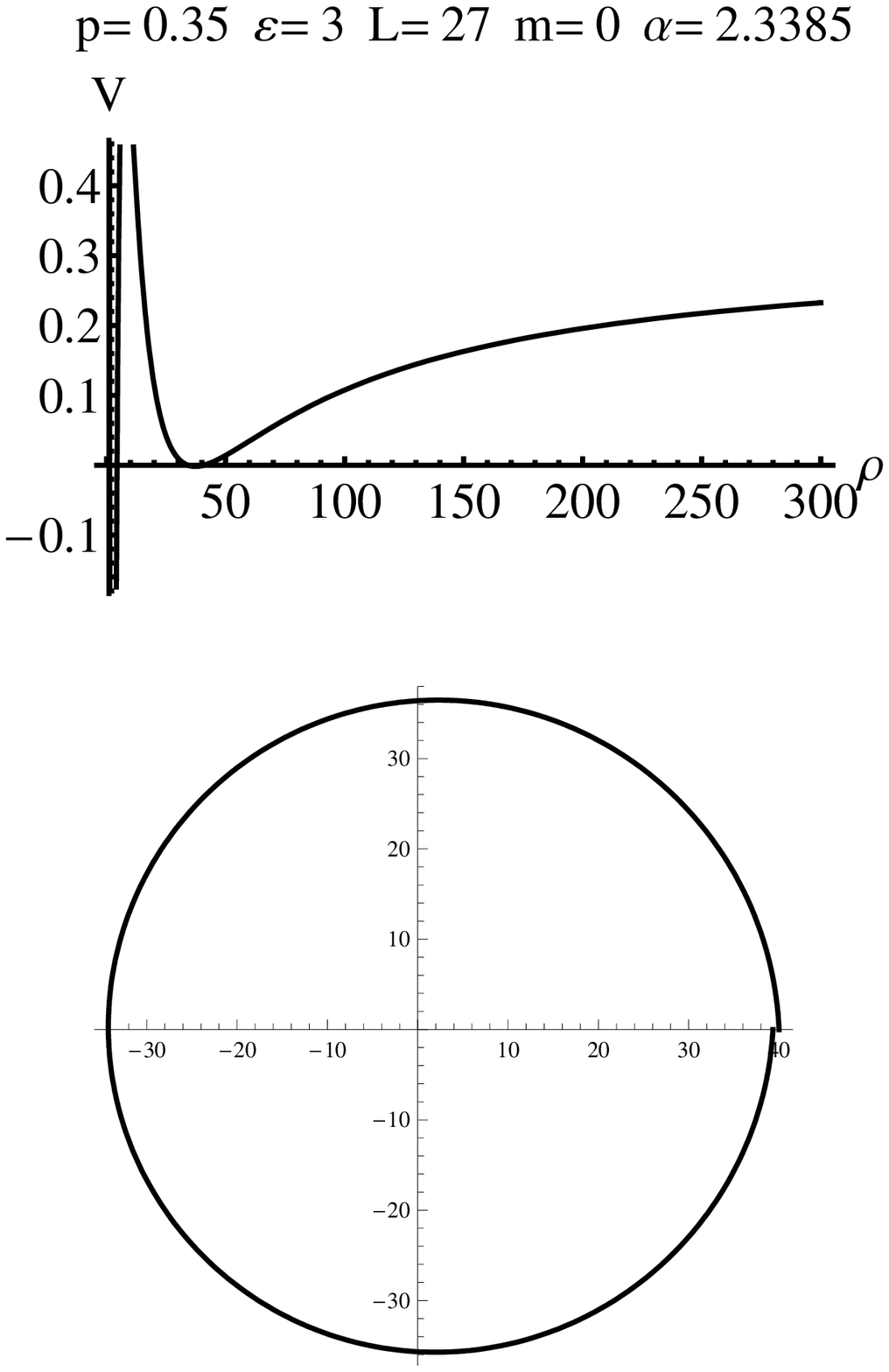}
\includegraphics[width=1.3in]{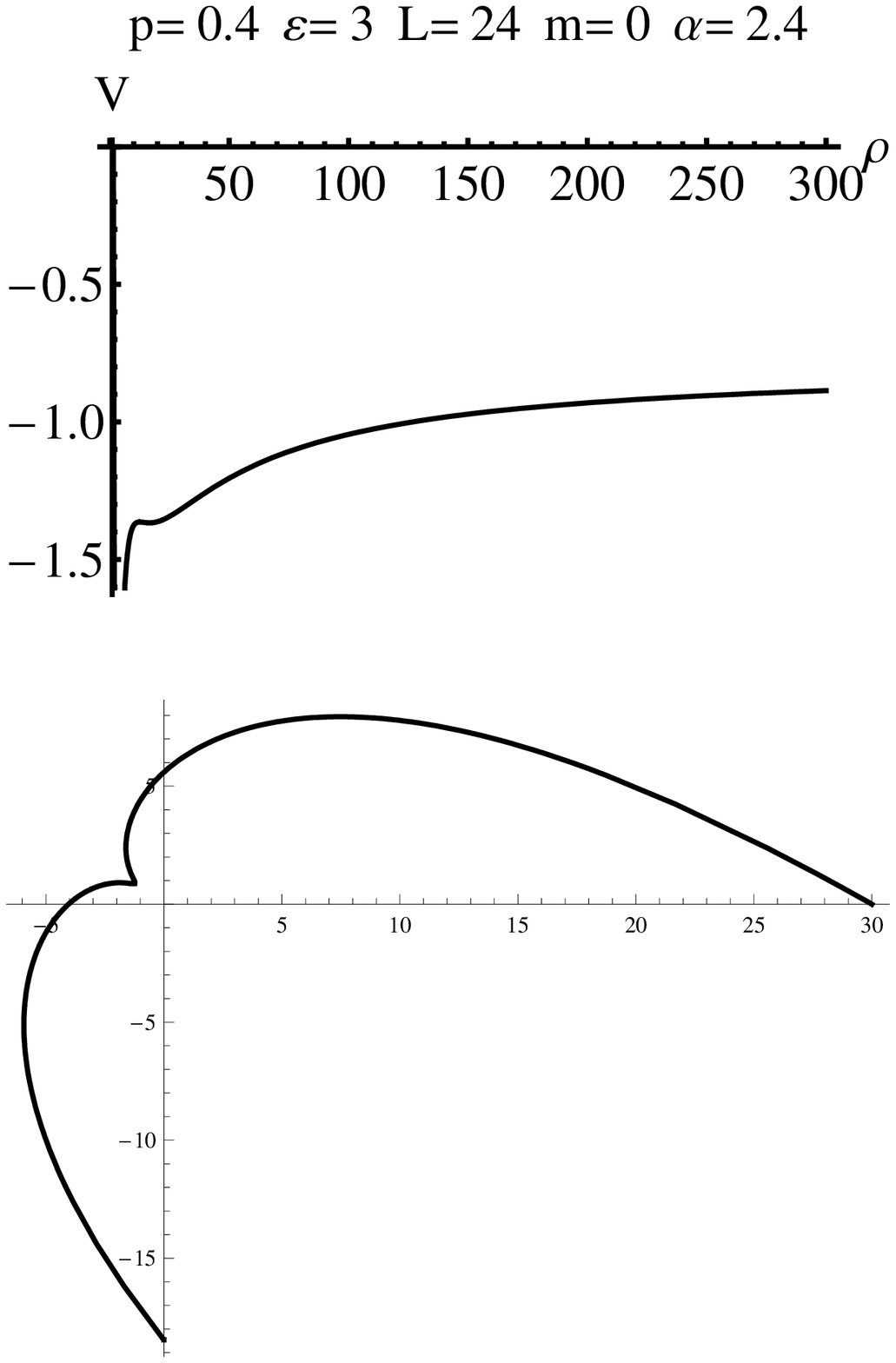}
\includegraphics[width=1.3in]{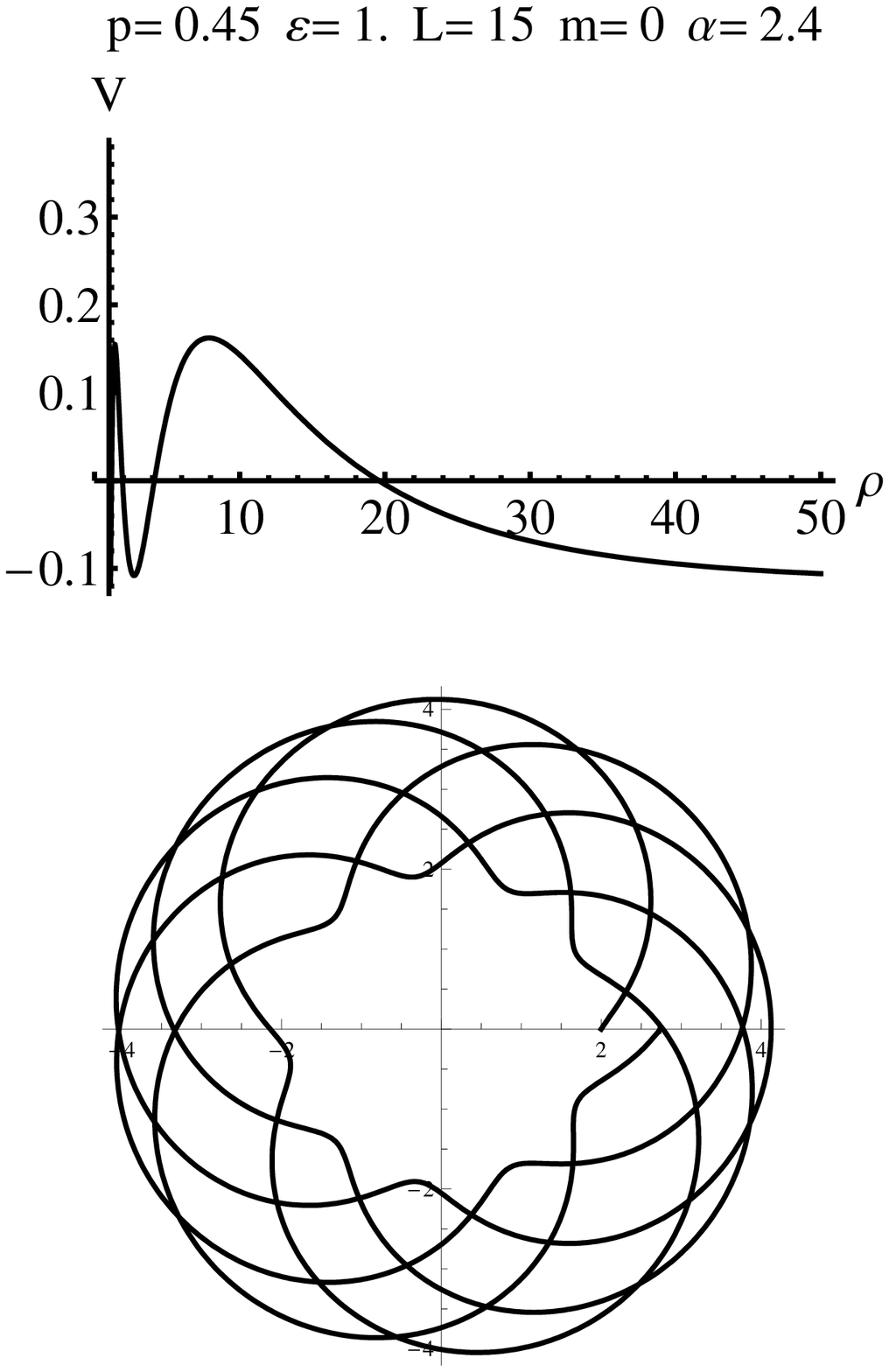}
\end{center}
\caption{\footnotesize{The effective potential and lightlike
geodesic orbits in black string spacetime with non-zero momentum along
the $z$-direction. }} \label{fig:fig9}
\end{figure}

\begin{figure}[h]
\begin{center}
\includegraphics[width=1.3in]{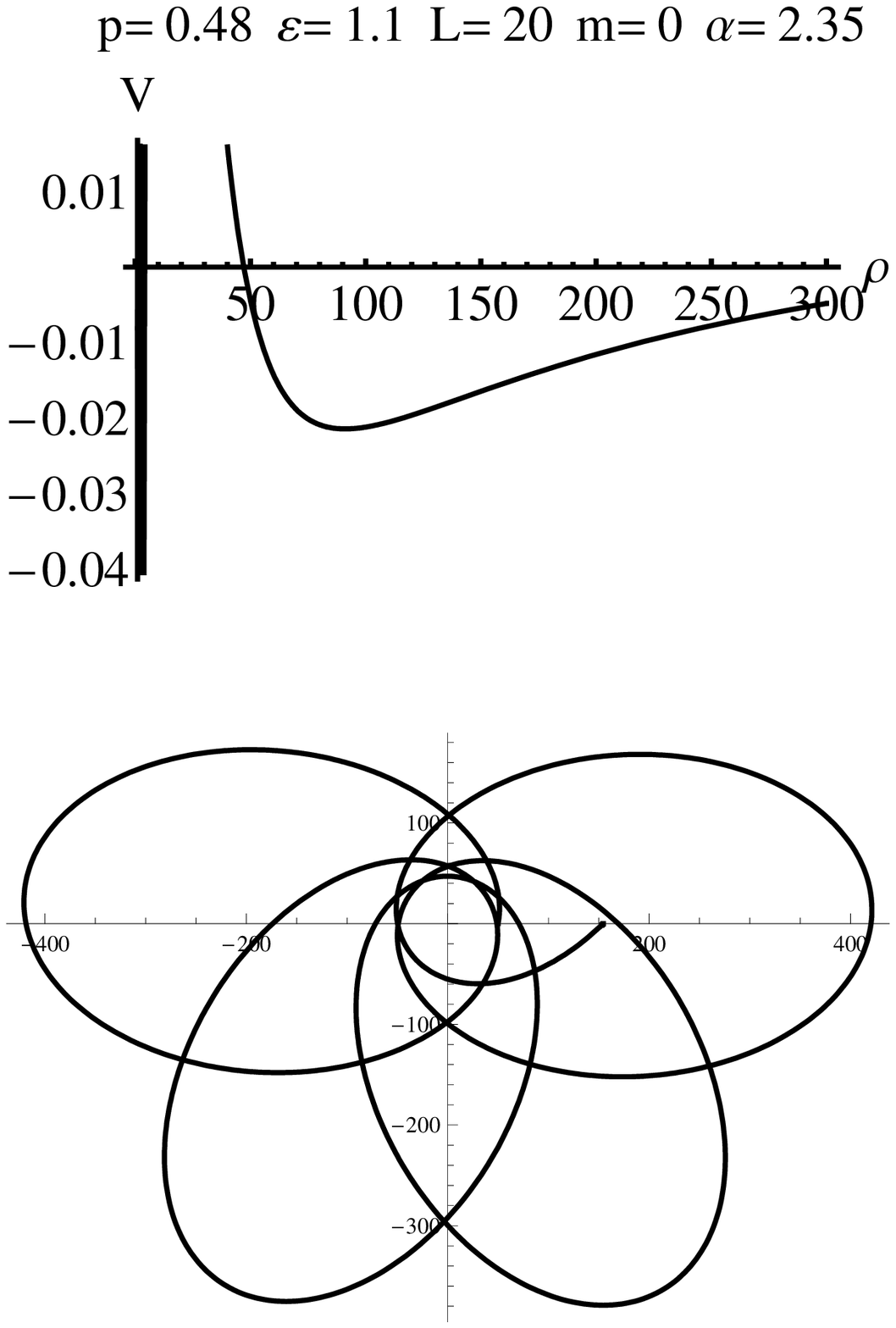}
\includegraphics[width=1.3in]{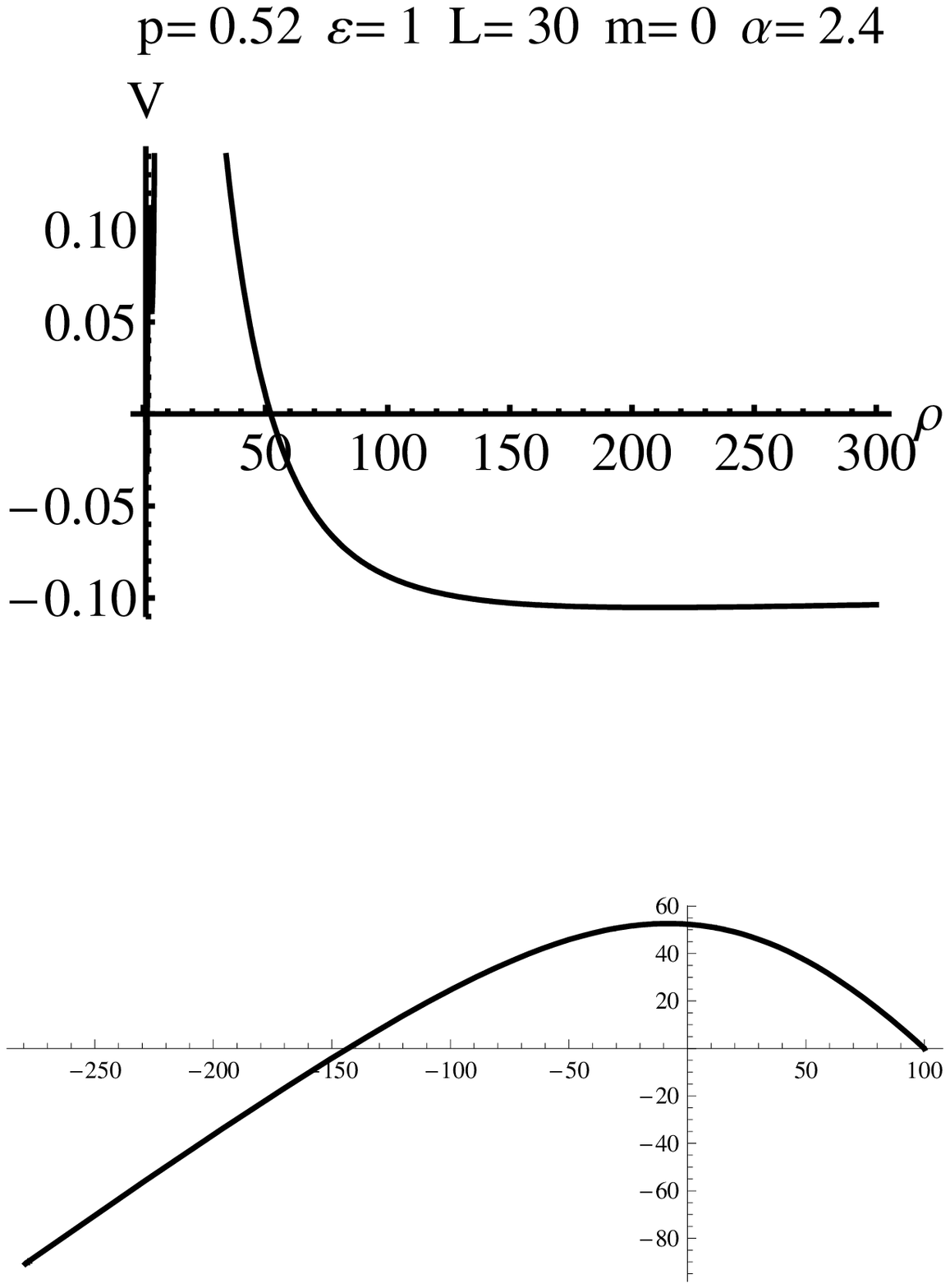}
\includegraphics[width=1.3in]{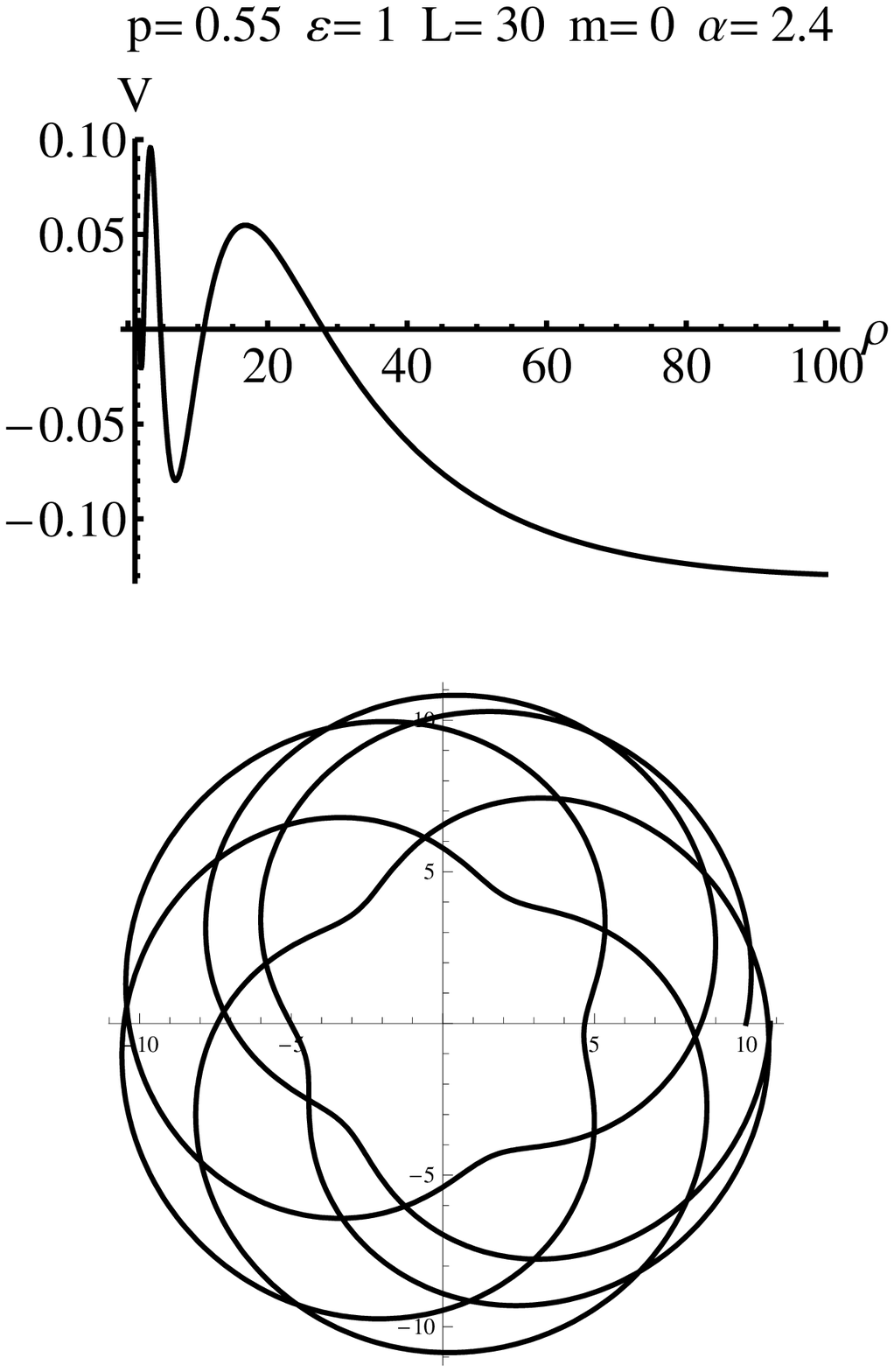}
\includegraphics[width=1.3in]{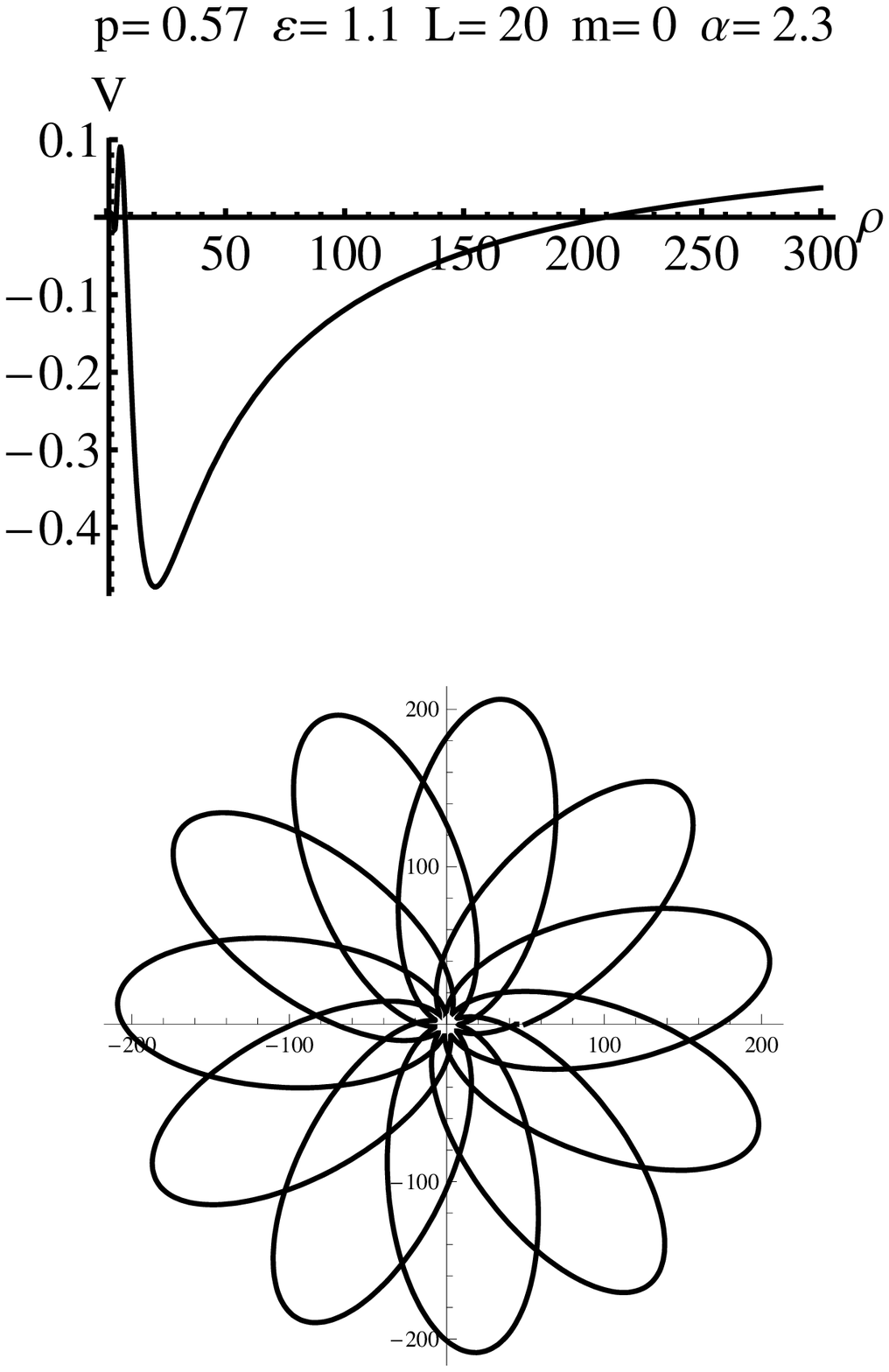}
\end{center}
\caption{\footnotesize{The effective potential and lightlike
geodesic orbits in wormhole geometry with non-zero momentum
along the $z$-direction. }} \label{fig:fig10}
\end{figure}

\begin{figure}[t]
\begin{center}
\includegraphics[width=4in]{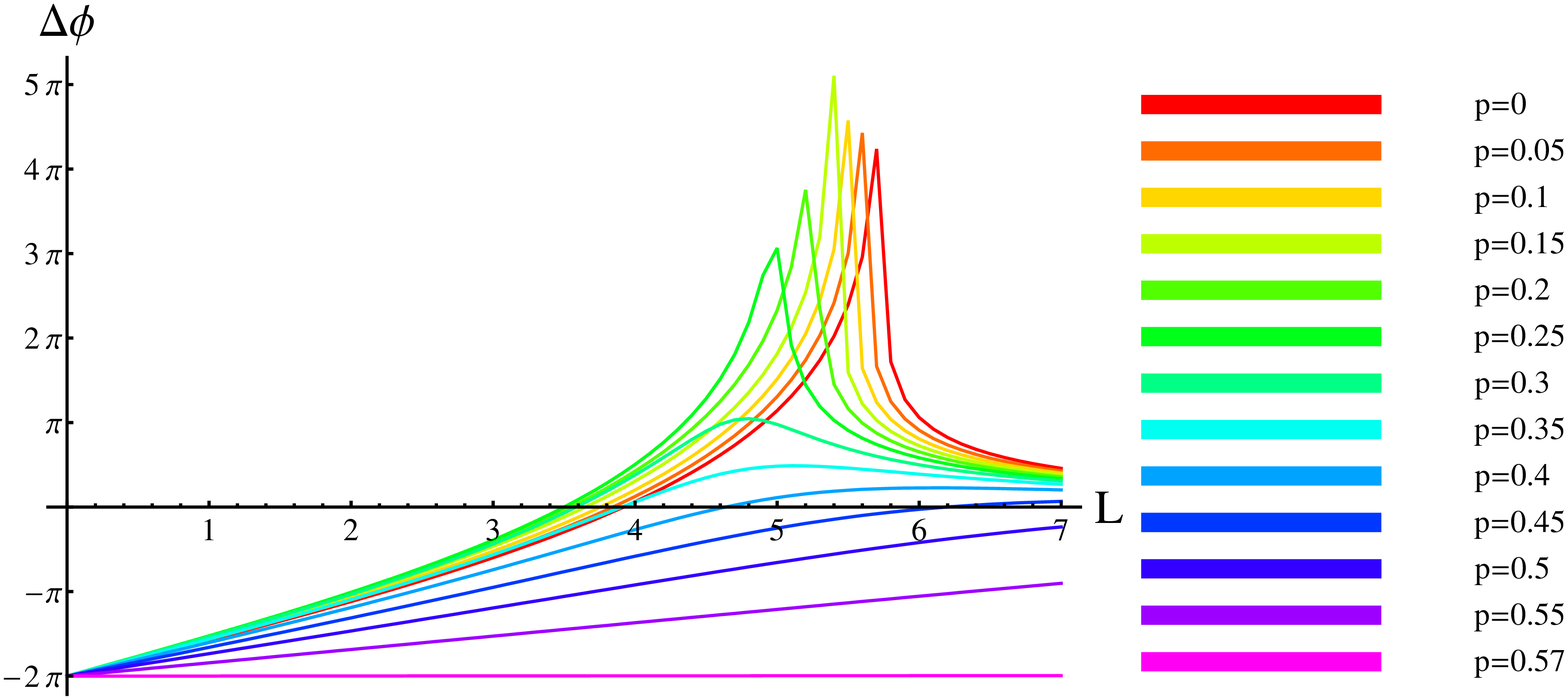}
\end{center}
\caption{\footnotesize{A plot of the precession rate as a function
of $\mathfrak{p}$, where $m=1$,
$\E=1$ and $\alpha=0.1$.}} \label{fig:fig11}
\end{figure}

In the present section, we explore the geodesic motions of a
massive particle and a light ray under the influence of the
effective potential (\ref{Veff}) in the extraordinary string
geometry. The geodesic motions of a massive particle and the
effective potentials for the corresponding motions in the plane at
$\theta=\pi/2$ are shown in Figs.\ \ref{fig:fig2}-\ref{fig:fig11}.
The corresponding parameter values are written in each figure. We
set $m=1$ for timelike geodesics. In this work, we divide the
timelike geodesics into two groups. The first group represents the
geodesics with non-zero momentum along the $z$-direction. The
second group represents the geodesics with zero momentum along the
$z$-direction. If we take the geometry with non-zero momentum along the $z$-direction, each figure shows that the curves are projected on the $X$-$Y$ plane. The particles move in three different geometries. The first geometry has the naked singularity. The second one corresponds to the black string geometry. The third one
corresponds to the wormhole geometry. In the analysis of the
geodesic motions, one can usually read off the behaviors of the
motions from the shape of the effective potential. However, we
visualize the trajectories in detail to show the unusual features
such as bouncing off the potential barrier at strong
gravitational regime. The particle resides only in the regions of
$V_{eff} \leq 0$.
In these figures, the radial coordinate $\rho$ is measured in unit
of $K$. For every trajectories, the potential barriers surrounding
the geometry exist always as is shown in each figure. If $V_{eff}$
has a local minimum or maximum with $V_{eff}=0$, there exists the
circular motion of a massive particle at that radius. In our
cases, there exist stable circular orbits at certain parameter values in each geometry. If $V_{eff}$ has the region of $V_{eff}<0$, there exist elliptic motions. The trajectories are bounded between perihelion and aphelion points. The second figure in Fig.~\ref{fig:fig2}, the third figure in Fig.~\ref{fig:fig3}, the first and the fourth figure in Fig.~\ref{fig:fig4}, and the third figure in Fig.~\ref{fig:fig7} reveal an unusual behavior showing the motion of a particle bouncing off the potential barrier. This behavior is qualitatively different from that of the spacetime without the ``momentum charge". The first figure in Fig.~\ref{fig:fig2}, the third figure in Fig.~\ref{fig:fig3}, the first figure in Fig.~\ref{fig:fig4}, the second figure in Fig.~\ref{fig:fig5}, and the third figure in Fig.~\ref{fig:fig6} indicate the hyperbolic motions.

Now, we study the lightlike geodesics ($m=0$) for non-zero
momentum along the $z$-direction. There exist a stable circular
motion (spiral motion if one include the motion along $z$), the elliptic motion,
the hyperbolic motion, and the
unusual behavior corresponding to the motion of a particle
bouncing off the potential barrier. There are no stable circular
orbits for light rays in the Schwarzschild case.
The lightlike geodesic motion can be used to understand the
gravitational lensing effect. The lensing is described by its
deflection angle. In Schwarzschild case, the effective deflection
angle of relativistic image is an important clue for getting
information of strong gravitational lensing \cite{Virb2}.

Next, we consider the massive particle bounded in an orbit around
the string solution. At perihelion and aphelion points,
$\rho$ reaches its minimum and maximum values $\rho_{min}$ and
$\rho_{max}$, where $d\rho/d\phi$ vanishes. The change in $\phi$
as $\rho$ decreases from $\rho_{max}$ to $\rho_{min}$ is the same
as the change in $\phi$ as $\rho$ increases, so the total change
in $\phi$ per revolution is $2[\phi(\rho_{max}) -
\phi(\rho_{min})]$. This is equal to $2\pi$ if the orbit was a
closed ellipse. Thus, the orbit precession angle is given by
\begin{equation}
\bigtriangleup \phi = 2\{[\phi(\rho_{max}) -
\phi(\rho_{min})]-\pi\},
\end{equation}
where $\phi(\rho_{max}) - \phi(\rho_{min})$ is determined by
Eq.~(\ref{eqphi}). If the orbit should precess in the same
direction as the motion of the particle, the rate has positive
value. In Fig.~\ref{fig:fig11} the curves represent the change of
the precession rate versus the values of $L$, where
$m=1$, $\E=1$ and $\alpha=0.1$. The ratio of the momentum charge-to-mass $\mathfrak{p}$ changes the precession rate. The rate $\bigtriangleup \phi$ is $-2\pi$ at $L=0$. This means that $\phi(\rho_{max}) - \phi(\rho_{min})=0$ in the present geometry. The numerical results show that the precession rate can be classified into three groups in the range of $\mathfrak{p}$.
The first group has the range from $0$ to about $0.3 \simeq 1/(2\sqrt{3})$, where $1/(2\sqrt{3})$ denotes the boundary of the naked singularity and black string. In this group, the curve has a peak. There is the highest peak about $\mathfrak{p}=0.15$. And then the magnitude of the peak is decreased as $\mathfrak{p}$ increases. The position of the peak is moved to the left as $\mathfrak{p}$ increases. The second group has the range from $0.3$ to $0.45$. The numerical results show that the curves have the tendency to converge to $-2\pi$ as the value of $\mathfrak{p}$ increases. Interestingly, the peak is moved to the right and gone away as $\mathfrak{p}$ increases.
As the angular momentum increases, there is a tendency to have a circular motion so that the precession angle $\Delta \phi =0$.
The third group has the ranges bigger than $0.5$, for that curves the maximum has gone away and it continually increases to zero with angular momentum.
Note that the boundary $\mathfrak{p}= \sqrt{2}/3$ between the black string and wormhole lies between the two values $0.45$ and $0.5$.
Noting these results, we can distinguish the properties of the geometry by analyzing the angular momentum dependence of the precession angle.

\section{Summary and discussions }

We studied the geodesic motions in extraordinary string solutions
in 4+1 dimensions. First we show that the extraordinary solution
developed in Ref.~\cite{kimjl} is classified by the mass and
``momentum charge" densities. Central properties of the solution
were summarized. Especially, the ``momentum charge"-to-mass ratio
is restricted to $|\mathfrak{p}|\leq 1/\sqrt{3}$. Its geometry
takes the form of a naked singularity, a black string with a null singularity, and a wormhole according to the value of
$|\mathfrak{p}|$.

The geodesic motion are obtained by solving the Hamilton-Jacobi equation to get the radial equation of motion. We analyzed the effective potential and showed that the geodesic motion may have many qualitatively different behaviors from those of  spacetimes without ``momentum charge".
Especially, any geodesic cannot arrive at $\rho=K$.
In order to arrive there, a particle needs to follow non-geodesic trajectory.
In addition, there are stable circular (or spiral) null geodesics which are absent for Schwarzschild geometry.
The asymptotic region is allowed to particles with energy larger than $\sqrt{p_z^2+m^2}$.
The existence condition is crucially dependent on $p_z$.

We explicitly presented some geodesic trajectories numerically.
In this work, we divided the timelike geodesics into two groups in terms of the value of momentum charge.
The particles move in three different geometries, spacetime with the naked
singularity, with the black string, or with the wormhole geometry.
In the analysis of the geodesic motions, one can usually read off the
behaviors of the radial motions from the shape of the effective
potential. We also study the lightlike geodesics ($m=0$) for
non-zero momentum along the $z$-direction. There exist stable
circular motions, elliptic motions, hyperbolic motions, and
the unusual behavior of a light ray bouncing off the potential barrier.

In addition, we illustrated the orbit precession. The precession angle
$\bigtriangleup \phi$ is $-2\pi$ at $L=0$ as expected.
The numerical results show that it can be classified three groups in terms of the range of $\mathfrak{p}$.
In addition, it is indicated that we can identify the near horizon (or singularity) $\rho\simeq K$ property of the solution by analyzing the angular momentum dependence of the precession angle since the precession angle shows distinct features for each geometry.
In the case of a naked singularity, as the angular momentum increases, the precession angle has a sharp peak which moves to the left as the ``momentum charge" increases.
In the case of a black string, it has a smooth peak which moves to the right.
In the case of a wormhole geometry, there is no peak and the angle continually increases and approaches to zero.

\begin{acknowledgments}
We would like to thank Gungwon Kang and Hee Il
Kim for their kind comments. This work was supported by the Korea
Science and Engineering Foundation (KOSEF) grant funded by the
Korea government(MEST) through the Center for Quantum Spacetime(CQUeST) of Sogang University with grant number R11-2005-021 and by the Korea Research Foundation Grant funded by the Korean Government(MOEHRD) KRF-2007-355-C00014(WL) and KRF-2008-314-C00063(H.-C. K.). BG is supported by the scholarship of Korea Science
and Engineering Foundation (Scholarship Number:S2-2008-000-00800-1).
\end{acknowledgments} \vspace{1cm}


\begin{thebibliography}{10}

\bibitem{GL}
R. Gregory and R. Laflamme, Phys. Rev. D {\bf 37}, 305 (1988).

\bibitem{after}
R. Gregory and R. Laflamme, Phys. Rev. Lett. {\bf 70}, 2837
(1993); T. Hirayama and G. Kang, Phys. Rev. D {\bf 64}, 064010 (2001).; S. S. Gubser and I. Mitra, High Energy Phys. {\bf 08}, 018 (2001).; M. W. Choptuik, L. Lehner, I. I. Olabarrieta, R. Petryk, F.
Pretorius, H. Villegas, Phys. Rev. D {\bf 68},  044001 (2003).; G.
Kang and J. Lee, J. High Energy Phys. {\bf 03}, 039 (2004); T. Harmark and N.
Obers, [hep-th/0503020]; H. Kudoh, Phys. Rev. D {\bf 73}, 104034
(2006); Y. Brihaye, T. Delsate, and E. Radu, Phys. Lett. B {\bf
662}, 264 (2008).

\bibitem{Horowitz}
G. T. Horowitz and K. Maeda, Phys. Rev. Lett. {\bf 87}, 131301
(2001); S. S. Gubser, Class. Quant. Grav. {\bf 19}, 4825 (2002).

\bibitem{Kramer}
D. Kramer, Acta Phys. Polon. B {\bf 2}, 807 (1971); J. Gross and
M. M. Perry, Nucl. Phys. {\bf B226}, 29 (1983); A. Davidson and D.
A. Owen, Phys. Lett. B {\bf 155}, 247 (1985).

\bibitem{chodos}
A. Chodos and S. Detweiler, Gen. Relativ. Gravit. {\bf 14} 879,
(1982).

\bibitem{lee}
C. H. Lee, Phys. Rev. D {\bf 74}, 104016 (2006).

\bibitem{ADM}
J. Traschen and D. Fox,
Class. Quant. Grav. {\bf 21}, 289 (2004)
[arXiv:gr-qc/0103106].

\bibitem{kang}
I. Cho, G. Kang, S. P. Kim and C. H. Lee; J. Korean Phys. Soc.
{\bf 53}, 1089 (2008);
G. Kang, H.-C. Kim, and J. Lee, Phys. Rev. D {\bf 79}, 124030
(2009).

\bibitem{yun} S.~Yun, Mod.\ Phys.\ Lett.\  A {\bf 25}, 159 (2010).

\bibitem{kimjl}
H.-C. Kim and J. Lee, Phys. Rev. D {\bf 77}, 024012 (2008).

\bibitem{Misner}
C. W. Misner, K. S. Thorne, and J. A. Wheeler, {\it Gravitation}
(Freeman, San Francisco, 1973).

\bibitem{matos}
T. Matos and N. Breton, Gen. Relativ. Gravit. {\bf 26} 827,
(1994).

\bibitem{gll}
B. Gwak, B.-H. Lee, and W. Lee, J. Korean Phys. Soc. {\bf 54},
2202 (2009).

\bibitem{jlkim}
J. Lee and H.-C. Kim, Mod. Phys. Lett. A {\bf 22}, 2439 (2007).

\bibitem{velocityFD}
J. Lee and H.-C. Kim, Mod. Phys. Lett. A {\bf 23}, 305 (2008).

\bibitem{Liu}
H. Y. Liu, P. Wesson and J. Poince de Leon, J. Math. Phys. {\bf
34}, 4070 (1993).

\bibitem{Billyard}
A. Billyard and P. S. Wesson, Phys. Rev. D {\bf 53}, 731 (1996).

\bibitem{ve}
K. S. Virbhadra and G. F. L. Ellis, Phys. Rev. D {\bf 65}, 103004
(2002).
%

\bibitem{Virb2}
  K.~S.~Virbhadra,
  Phys.\ Rev.\  D {\bf 79}, 083004 (2009).



\end{thebibliography}

\vspace{4cm}

\end{document}